\documentclass[11pt]{article}
\pdfoutput=1  
\usepackage{graphicx}
\usepackage{subfig}
\graphicspath{{./figures/}}
\usepackage{appendix}
\usepackage{latexsym,amsmath,amsfonts,amssymb,booktabs}
\usepackage[font=small]{caption}
\usepackage{slashed,upgreek,amscd,cancel,tensor,color}
\usepackage{adjustbox}
\usepackage[numbers,compress,square]{natbib}
\usepackage{epsfig,latexsym}
\usepackage[pdfencoding=auto]{hyperref}
\usepackage{url}
\numberwithin{equation}{section}
\usepackage{doi}
\definecolor{MyBlue}{rgb}{0.15,0.15,0.70}

\hypersetup{
colorlinks=true,
citecolor=MyBlue,
linkcolor=MyBlue,
urlcolor=MyBlue
}

\input epsf
\usepackage{amssymb}
\usepackage{amsmath}
\usepackage{amsfonts}
\usepackage{upgreek}
\usepackage{latexsym}
\usepackage{stfloats}
\usepackage{afterpage}

\newcommand{\bfv}{\mbox{\boldmath$v$}}
\newcommand{\bfx}{\mbox{\boldmath$x$}}
\newcommand{\bfk}{\mbox{\boldmath$k$}}

\newcommand{\bfq}{\mbox{\boldmath$q$}}

\newcommand{\bfs}{\mbox{\boldmath$s$}}

\newcommand{\knl}{k_{\rm NL}}
\newcommand{\eqn}[1]{eq.~(\ref{#1})}
\newcommand{\eqnstwo}[2]{eqs.~(\ref{#1})~and~(\ref{#2})}

\newcommand{\be}{\begin{equation}}
\newcommand{\ee}{\end{equation}}
\newcommand{\sect}[1]{Sec.~\ref{#1}}
\newcommand{\secref}[1]{Sec.~\ref{#1}}
\newcommand{\citee}[1]{\cite{#1}}
\newcommand{\citees}[1]{\cite{#1}}
\newcommand{\appref}[1]{App.~\ref{#1}}
\newcommand{\figref}[1]{Fig.~\ref{#1}}
\newcommand{\figrefs}[2]{Figs.~\ref{#1} and \ref{#2}}
\newcommand{\figrefss}[3]{Figs.~\ref{#1},~\ref{#2}, and \ref{#3}}

\newcommand{\tabref}[1]{Tab.~\ref{#1}}
\newcommand{\ta}{\tilde a}

\newcommand{\omegam}{\Omega_{m}}
\newcommand{\cH}{\mathcal{H}}
 \newcommand{\momspmeas}[1]{\frac{d^3 #1}{(2 \pi)^3}}
 \newcommand{\kvec}{\bfk}
 \newcommand{\qvec}{\bfq}
 
 \newcommand{\kdotq}{\hat k \cdot \hat q}
 \newcommand{\bea}{\begin{array}}
 \newcommand{\eea}{\end{array}}
\newcommand{\unitsk}{\, h { \rm Mpc^{-1}}}
\newcommand{\omegarc}{\Omega_{\rm rc}}
\newcommand{\kfit}{k_{\rm fit}}
\newcommand{\kreach}{k_{\rm reach}}

\begin{document}
\vspace{0.5cm}

\begin{center}
\Large{\textbf{Towards Precision Constraints on Gravity \\ with the Effective Field Theory of Large-Scale Structure }} \\[1cm]

\large{Benjamin Bose$^{1}$, Kazuya Koyama$^{1}$, Matthew Lewandowski$^{2}$, \\ Filippo Vernizzi$^{2}$, Hans A. Winther$^{1}$}
\\[0.5cm]

\small{
\textit{$^1$Institute of Cosmology \& Gravitation, University of Portsmouth,
Portsmouth, Hampshire, PO1 3FX, UK}}

\vspace{.2cm}

\small{
\textit{$^2$ Institut de physique th\' eorique, Universit\'e  Paris Saclay 
CEA, CNRS, 91191 Gif-sur-Yvette, France }}

\vspace{.2cm}

\vspace{0.5cm}
\today

\end{center}

\vspace{2cm}

\begin{abstract}

 We compare analytical computations with numerical simulations for dark-matter clustering, in general relativity and in the normal branch of DGP gravity (nDGP).  
Our analytical frameword is the Effective Field Theory of Large-Scale Structure (EFTofLSS), which we use to compute the one-loop dark-matter power spectrum, including the resummation of infrared bulk displacement effects. We compare this to a set of  20 COLA  simulations at redshifts $z = 0$, $z=0.5$, and $z =1$, and fit  the free parameter of the EFTofLSS, called the speed of sound, in both $\Lambda$CDM and nDGP at each redshift.  
At one-loop at $z = 0$, the reach of the EFTofLSS is $\kreach \approx 0.14 \unitsk$ for both $\Lambda$CDM and nDGP.
Along the way, we compare two different infrared resummation schemes and two different treatments of the time dependence of the perturbative expansion, concluding that they agree to approximately $1\%$ over the scales of interest. 
Finally, we use the ratio of the COLA power spectra to make a precision measurement of the difference between the speeds of sound in $\Lambda$CDM and nDGP, and verify that this is proportional to  the modification of the linear coupling constant of the Poisson equation.

\end{abstract}
\hspace{.35in} PACS numbers: 98.80.-k

\newpage

\tableofcontents

\vspace{.5cm}
\newpage

\section{Introduction}


 Despite the success of the $\Lambda$CDM model, 
alternatives, coming in the form of modifications to gravity or  to the energy components of the universe, are currently under active investigation (see \citee{Bull:2015stt} for a review).
Generically, modified gravity theories predict a fifth force sourced by additional degrees of freedom. To overcome strong local experimental constraints on gravity, all largely consistent with GR, modified gravity theories must employ screening mechanisms. These mechanisms filter out fifth force effects at small scales (see \citees{Koyama:2015vza,Sakstein:2015oqa,Clifton:2011jh} for reviews) so that the theory does not violate the strong experimental tests.  These short-scales constraints have pushed searches for deviations from GR to the  large-scale structure of the universe \citees{Uzan:2000mz,Lue:2004rj,Ishak:2005zs,Knox:2005rg,Koyama:2006ef,Chiba:2007rb,Amendola:2007rr,Simpson:2012ra,Terukina:2012ji,Terukina:2013eqa,Yamamoto:2010ie,Jain:2007yk,Zhao:2008bn,Zhao:2009fn,Asaba:2013xql}.  On the observational side, this field will soon enter an exciting new era with the commencement of the largest and most precise astronomical surveys to date, including EUCLID\footnote{\url{www.euclid-ec.org}} \citee{Laureijs:2011gra}, WFIRST\footnote{\url{https://wfirst.gsfc.nasa.gov/}} \citee{Spergel:2013tha}, DESI\footnote{\url{http://desi.lbl.gov/}}  \cite{Aghamousa:2016zmz}, and LSST\footnote{\url{https://www.lsst.org/}} \citee{Chang:2013xja}. 

One widely used quantity in survey data analyses is the matter power spectrum, used both in lensing and in clustering experiments \citees{Crocce:2012fa,Taruya:2007xy,Pietroni:2008jx,McDonald:2006hf,Matsubara:2007wj,Bernardeau:2011dp,Blas:2015qsi,Kilbinger:2012qz,Schrabback:2009ba}. Currently, perturbative templates are widely used to compute this quantity (see \citees{Bernardeau:2001qr,Bartelmann:1999yn} for early reviews of relevant theory). In order to make the most of the upcoming data sets, such approaches must improve on two fronts.  Firstly, because most information is concentrated at short scales, one would like to be able to accurately describe the mildly non-linear regime.  Secondly, as the sizes of the surveys increase, the statistical errors are pushed into the percent and sub-percent regime: this calls for precise control over theoretical errors and for common approximations used in theoretical templates to be quantified and, if necessary, improved. 

The first of these problems has recently been tackled using the effective field theory of large-scale structure (EFTofLSS) approach \citees{Baumann:2010tm,Carrasco:2012cv,Porto:2013qua,Mercolli:2013bsa,Carrasco:2013mua,Senatore:2014via,Baldauf:2014qfa,Lewandowski:2014rca,Assassi:2015jqa,Foreman:2015uva,Baldauf:2015xfa,Foreman:2015lca,Bertolini:2015fya,Bertolini:2016bmt,Bertolini:2016hxg,Perko:2016puo,Cusin:2017wjg}.  After it was noted that including higher orders in the standard perturbation theory (SPT) expansion did not seem to ensure a better modeling of the non-linear regime \citee{Carlson:2009it}, the EFTofLSS was developed to provide a consistent, controllable expansion.  In order to correctly describe gravitational clustering at the smallest scales possible, the EFTofLSS correctly treats, in perturbation theory, the effects of small-scale (i.e ultraviolet, or UV) modes on the long-wavelength (i.e. infrared, or IR) observables measured in LSS surveys.  

The main idea of this approach is to include, in the dark-matter equations of motion, all operators which are consistent with the equivalence principle and other underlying symmetries.  These additional terms can be systematically organized in terms of powers of fields and number of derivatives, so that for a given desired precision in some computation, only a finite number of the operators needs to be retained.  These additional terms, which come with free coefficients not predictable within the EFT, systematically correct mistakes introduced in loops from uncontrolled short-distance physics.  These counterterms are essential for a consistent and accurate description of clustering; if we are to place trusted constraints on gravity in the context of stage IV surveys, such an approach should be well considered.  

The second problem, that of limitations of commonly used approximations, techniques, and templates used in comparisons of data to theory, has been investigated in a number of works \citees{Taruya:2010mx,Bose:2016qun,Barreira:2016mg,Taruya:2016jdt,Bose:2017myh,Lewandowski:2017kes}, many in the context of modeling modifications of gravity, and we continue that investigation here.  In this work we aim to provide comparisons and quantify the effects of various theoretical approximations on the matter power spectrum in the EFTofLSS framework. In particular, we look at the effects of two different approaches to solve for the time dependence and two different approaches to resum IR displacement effects (which is necessary to correctly describe the baryon acoustic oscillations (BAO) in the power spectrum, see for example \citees{Matsubara:2007wj,Senatore:2014via,Baldauf:2015xfa,Vlah:2015sea,Vlah:2015zda}).  We find that, in general, the differences between these approaches are sub-percent.  We also investigate the dependence of the sound speed parameter of the EFTofLSS on modifications to gravity.  To do this, we consider two gravitational models, GR and the braneworld model of gravity by Dvali, Gabadadze, and Porrati  \citee{Dvali:2000hr}. In particular, we look at the normal branch, called nDGP.  In this work, we concentrate on nDGP because we have available to us the measurements from $20$ COmoving Lagrangian Acceleration (COLA) simulations, each of size $1\mbox{Gpc}^3 \, h^{-3}$ \citee{Winther:2017jof}.  Although here we focus on a specific model of modified gravity, our methods can be applied to other models of modified gravity.  For instance, it can be straightforwardly applied to Horndeski class of theories in the effective field theory of dark energy (EFTofDE) approach \cite{Gubitosi:2012hu,Gleyzes:2013ooa,Gleyzes:2014rba} developed in \citee{Cusin:2017mzw,Cusin:2017wjg}.  

This paper is organized as follows.  In \sect{theorysec} we review the one-loop computations in SPT and the EFTofLSS, using both exact and EdS (Einstein de Sitter) time dependence, and we also review two different IR-resummation techniques. In \sect{resultssec} we compare the predictions at $ z = 0, 0.5$, and $ 1$ using different approximations and resummation schemes within GR and nDGP, and determine an approximate validity range of the theoretical frameworks.  
We also investigate the dependence of the EFTofLSS counterterm on the nDGP parameter which determines the strength of the modification of gravity.  Finally in \sect{concsec} we summarize our results and highlight future work.

%
%

%
%

\section{Theoretical Framework} \label{theorysec} 
\subsection{Setup in standard perturbation theory}
\label{sub_setup}

 In DGP gravity \citee{Dvali:2000hr} we live on a four-dimensional brane embedded in five-dimensional Minkowski spacetime. This produces a crossover scale $r_c$ (the only free parameter in the theory), which is given by the ratio between the five-dimensional Newton's gravitational constant and the four-dimensional Newton's  gravitational constant.   The modified Friedman equation is given by 
\begin{equation}
\epsilon\frac{H}{r_c} = H^2 \big(1 - \Omega_m(a) \big) \ , 
\end{equation}
where $H \equiv \dot a/a$ is the Hubble rate, $\Omega_m \equiv 8 \pi G \rho_m/(3 H^2)$ is the fractional dark matter energy density and $\epsilon=\pm 1$. 

The solution for $\epsilon=1$ was found to be unstable \cite{Luty:2003vm}. We thus consider the stable normal branch with $\epsilon = -1$ (nDGP). In this branch, acceleration is achieved through a cosmological constant as in GR. To simplify the analysis, we impose a background history following $\Lambda$CDM, done by tuning the dark energy equation of state. Therefore, we have
\be
H(a ) = H_0 \sqrt{ \Omega_{ m0} a^{-3}  + \left( 1 - \Omega_{ m  0} \right) } \ , \hspace{.2in} \text{and} \hspace{.2in} \Omega_{ m} (a) = \frac{ \Omega_{ m0} a^{-3} }{  \Omega_{ m 0} a^{-3} + (1 - \Omega_{ m 0})   } \;,
\ee
where $H_0$ is the Hubble parameter today

We consider scalar perturbations around the FRLW metric, which in Newtonian gauge can be written as
\begin{equation}
ds^2=-(1+2\Phi)dt^2+a(t)^2(1-2\Psi)\delta_{ij}dx^idx^j.
\end{equation}
The Newtonian potential $\Phi$  is governed by the modified Poisson equation \citee{Koyama:2009me}. In Fourier space, this reads 
\begin{equation}
-\left(\frac{k}{a H(a)}\right)^2\Phi (\bfk;a)=
\frac{3 \Omega_m(a)}{2} \mu(a)\,\delta(\bfk;a) + S(\bfk;a),
\label{eq:poisson1}
\end{equation}
where $\delta$ is the dark matter density contrast.  The function $\mu( a )$ (which for general modified gravity models can depend also on the scale $k$) characterizes the linear modifications to the clustering equations and is given by
\begin{equation}
\mu \equiv 1 + \frac{1}{3\beta}  \ , \qquad 
 \label{betadef}
\beta \equiv 1+\frac{H}{H_0} \frac{1}{\sqrt{\Omega_{rc}}}   \left(1+\frac{aH'}{3H}\right) \ ,
\end{equation}
where  the prime denotes a derivative with respect to the scale factor $a$ and $H'=dH/da$. Here we choose to parameterize the cross-over scale in terms of $\Omega_{rc} \equiv 1/(4r_c^2H_0^2)$.

The non-linear source term $S(\bfk ; a )$ characterizes new mode couplings, including those responsible for screening effects. Up to third order in the perturbations, for nDGP this reads \citee{Bose:2016qun}
\begin{align}
S(\bfk; a) = \ &
  \mu_{ 2}     \left( \frac{ 3 \, \omegam}{2  } \right)^2    \int\frac{d^3\bfk_1d^3\bfk_2}{(2\pi)^3}\,
\delta_{\rm D}(\bfk-\bfk_{12}) \gamma_2( \bfk_1, \bfk_2)
\delta(\bfk_1)\,\delta(\bfk_2)
\label{eq:Perturb3} \\
& +  \mu_{22}    \left( \frac{ 3 \, \omegam }{2  } \right)^3 
\int\frac{d^3\bfk_1d^3\bfk_2d^3\bfk_3}{(2\pi)^6}
\delta_{\rm D}(\bfk-\bfk_{1 2 3})
\gamma_2 ( \kvec_2 , \kvec_3  )  \gamma_2 ( \kvec_1 , \kvec_2 + \kvec_3) 
\delta(\bfk_1)\,\delta(\bfk_2)\,\delta(\bfk_3) \ ,
\nonumber
\end{align}
with $\bfk_{1 \ldots n} \equiv \bfk_{1} + \ldots + \bfk_{n}$,
\begin{equation}
\gamma_2 ( \bfk_1 , \bfk_2 ) \equiv  1 - \frac{(\kvec_1 \cdot \kvec_2)^2}{k_1^2 k_2^2}  \;, 
\end{equation}
and
\begin{equation} \label{gamma2}
\mu_{ 2}  = - 2  H^2 r_c^2  \left( \frac{1}{3 \beta} \right)^3 \;, \qquad \qquad \mu_{22} = 8 H^4 r_c^4 \left( \frac{1}{3 \beta} \right)^5\;.
\end{equation}
General relativity is recovered when $\mu = 1$ and $\mu_2 = \mu_{22} = 0$.

Note that the same structure for the above non-linear corrections also appears  in the context of the nonlinear  EFTofDE \citee{Cusin:2017mzw,Cusin:2017wjg}, although DGP model cannot be described by the ordinary EFTofDE action because it is not a local theory of a scalar field that restores four-dimensional time diffeomorphisms.
However, because the form of the modified Poisson equation is the same, one can easily adapt the computational strategy used in \citee{Cusin:2017wjg} to the case presented here.\footnote{The nonlinear equations for the metric potentials $\Phi$ and $\Psi$ and scalar field perturbations $\varphi$ derived in Ref.~\cite{Koyama:2007ih} can be obtained from the nonlinear equations (3.22) of Ref.~\cite{Cusin:2017wjg}, with the following
 (non-unique) replacements:
$ \alpha_{\rm M}  = 2 \alpha_{\rm B} $,  $ \nu  = {3 \beta}/{\alpha_{\rm B}^2}  $,   $ \mathcal{C}_4 = 8 H^2 r_c^2 \alpha_{\rm B}^3 $, $\alpha_{\rm T} = \mathcal{C}_3 = \mathcal{C}_5 = 0$, and $   \chi= - \varphi/ (2 \alpha_{\rm B}   )   $.}
  

Let us now turn to the dark matter description.
The evolution equations for dark matter perturbations are obtained from the conservation of the matter energy momentum tensor.  First, we will present the equations relevant for SPT, i.e. neglecting counterterms from the EFTofLSS.  We discuss the modifications from the EFTofLSS in \sect{eftsec}.  Before shell crossing, neglecting vorticity in the velocity field\footnote{This is a safe assumption at large scales and late times, since vorticity is only generated at higher order in perturbation theory \citee{Carrasco:2013mua}.} and assuming the dark-energy field only interacts with matter through gravity, the evolution equations can be expressed in Fourier space as 
\begin{align}
&a \frac{\partial \delta(\bfk;a)}{\partial a}+\theta(\bfk;a) =-
\int\frac{d^3\bfk_1d^3\bfk_2}{(2\pi)^3}\delta_{\rm D}(\bfk-\bfk_{12})
\alpha(\bfk_1,\bfk_2)\,\theta(\bfk_1)\delta(\bfk_2) \ ,
\label{eq:Perturb1}\\
& a \frac{\partial \theta(\bfk;a)}{\partial a}+
\left(2+\frac{a H'}{H}\right)\theta(\bfk;a)
-\left(\frac{k}{a\,H}\right)^2\,\Phi(\bfk;a)=
-\int\frac{d^3\bfk_1d^3\bfk_2}{(2\pi)^3}
\delta_{\rm D}(\bfk-\bfk_{12})
\beta(\bfk_1,\bfk_2)\,\theta(\bfk_1)\theta(\bfk_2)  \ ,
\label{eq:Perturb2}
\end{align}
where  $\theta$ is the velocity divergence expressed in terms of the peculiar velocity field $\bfv_p({\bfx})$ as $\theta({\bfx})= \frac{\nabla\cdot \bfv_p({\bfx})}{aH(a)}$. The kernels in the Fourier integrals, $\alpha$  and $\beta$, are given by 
\begin{eqnarray}
\alpha(\bfk_1,\bfk_2)=1+\frac{\bfk_1\cdot\bfk_2}{|\bfk_1|^2},
\quad\quad
\beta(\bfk_1,\bfk_2)=
\frac{(\bfk_1\cdot\bfk_2)\left|\bfk_1+\bfk_2\right|^2}{2|\bfk_1|^2|\bfk_2|^2}.
\label{alphabeta}
\end{eqnarray}

%
%
%
\subsection{Linear solutions, SPT expansion, and Green's functions} \label{linearsolsec}

We would like to solve \eqn{eq:Perturb1} and \eqn{eq:Perturb2} perturbatively in the linear solution $\delta_1 ( \bfk;a)$, so we expand
\be
\delta ( \bfk ; a ) = \delta_1 ( \bfk ; a ) + \delta_2 ( \bfk ; a ) + \delta_3 ( \bfk ; a) + \dots   \hspace{.1in} \text{and} \hspace{.1in} \theta ( \bfk ; a ) = \theta_1 ( \bfk ; a ) + \theta_2 ( \bfk ; a ) + \theta_3 ( \bfk ; a) + \dots \ , 
\ee
where schematically $\delta_n \sim [ \delta_1]^n$.  
The linear equation for $\delta_1 ( \bfk;a)$ is given by 
\be  \label{lineaeq}
a^2  \delta_1 ( \bfk ; a )'' + a \left( 3 + \frac{a H' }{H} \right) \delta_1 ( \bfk ; a )'  - \mu(  a ) \frac{ 3 \, \Omega_m ( a ) }{2} \delta_1 ( \bfk ; a ) = 0 \ .  
\ee
For the DGP model that we consider here, $\mu(  a )$ does not depend on $k$, so the linear equation is scale-free.\footnote{ In general, when the description of gravity changes one must solve for the $\kvec$ and $a$ dependence simultaneously in a partial differential equation  \citees{Bernardeau:1993qu,Bouchet:1992uh,Catelan:1994kt,Bose:2016qun,Fasiello:2016qpn}.}  The solution is given by a growing mode, called $D_+(a)$, and a decaying mode $D_-(a)$.  Focusing on the growing mode at late times, this means we can write the linear solution as 
\be 
\delta_1 ( \bfk ; a ) = \frac{D_+(a) }{D_+(a_i )} \delta_1( \bfk ; a_i) \hspace{.4in} \text{and} \hspace{.4in} \theta_1 ( \bfk ; a ) =- \frac{ a D_+(a)'}{D_+(a_i)} \delta_1 ( \bfk ; a_i )  \ , \label{lin_sol} 
\ee
where $a_i$ is the time at which we set initial conditions.

These linear solutions then source non-linear corrections in eq.~(\ref{eq:Perturb1}) and eq.~(\ref{eq:Perturb2}), which we can write generically as (see \citee{Bernardeau:2001qr} for a review in the context of $\Lambda$CDM) 
\begin{align}  
\delta_n(\boldsymbol{k} ; a) &= \int d^3\boldsymbol{k}_1...d^3 \boldsymbol{k}_n \delta_D(\boldsymbol{k}-\boldsymbol{k}_{1...n}) F_n(\boldsymbol{k}_1,...,\boldsymbol{k}_n ; a) \delta_i(\boldsymbol{k}_1)...\delta_i(\boldsymbol{k}_n), \label{nth1} 
\end{align}
where $\boldsymbol{k}_{1...n} = \boldsymbol{k}_1 + ...+ \boldsymbol{k}_n$, $F_n$ is the \emph{n}-th order kernel, and $\delta_i(\bfk) = \delta_1(\bfk, a_i)$ is the field evaluated at the initial time $a_i$.  In order to compute the one-loop power spectrum, we will expand $\delta$ up to third order.

In the following, we consider two different methods of solving for the time dependence in \eqn{nth1}: the first is the use of the exact Green's functions, and the second is a hybrid between the EdS approximation and solving the exact time dependence.

First, we review the scheme that uses the exact Green's functions, which has also been employed in the context of the Horndeski dark-energy models in \citee{Takushima:2013foa,Takushima:2015iha,Cusin:2017wjg}.  Using the two linear solutions $D_+(a)$ and $D_-(a)$, one can construct the four Green's functions for the system \eqn{eq:Perturb1} and \eqn{eq:Perturb2},  $G^\delta_1 (a, \tilde a)$, $G^\delta_2(a, \tilde a)$, $G^\Theta_1(a, \tilde a)$ and $G^\Theta_2(a, \tilde a)$, which are explicitly given in \appref{exacttimeapp}. The Green's function $G^\delta_1$ gives the response of $\delta$ to a perturbation to the continuity equation, $G^\delta_2$ gives the response of $\delta$ to a perturbation to the Euler equation, and similarly for $\theta$.
Then, the perturbative solutions of the system can  be written as\footnote{The notation used in this paper is slightly different than that of \citee{Cusin:2017wjg}, where the authors worked with $\Theta \equiv - \theta$ and multiplied both sides of \eqn{eq:Perturb2} by $-1$.  Because we would like the equations of \citee{Cusin:2017wjg} to be directly applicable to our work here, \eqnstwo{dtGreen}{dtGreen2} look slightly unnatural because $S_n^{(2)}$ is the negative of the source term in \citee{Cusin:2017wjg}.}
 \begin{align}  \label{dtGreen}
 &\delta_n ( \bfk , a) =\int^a_0 d\ta \bigg(G^{\delta}_{1}(a,\ta)S^{(1)}_n(\bfk, \ta) - G^{\delta}_{2}(a,\ta)S^{(2)}_n(\bfk, \ta)\bigg) \ ,\\
  &\theta_n (\bfk , a )= - \int^a_0 d\ta \bigg(G^{\Theta}_{1}(a,\ta)S^{(1)}_n(\bfk, \ta) - G^{\Theta}_{2}(a,\ta)S^{(2)}_n (\bfk, \ta)\bigg) \ , \label{dtGreen2}
  \end{align}
where  the source terms $S^{(i)}_n$ are the $n$-th order expansion of the right-hand sides of \eqn{eq:Perturb1} for $i=1$, and \eqn{eq:Perturb2} for $i=2$, after plugging in the modified Poisson equation \eqn{eq:poisson1}.  In general, the $n$-th order source term is proportional to $n$ powers of the linear field, i.e. $S^{(i)}_n \sim [ \delta_1]^n$, and the $k$ dependence is dictated by the particular dependence of the non-linear vertices.  For example, one has 
\be
F_2 ( \kvec_1 , \kvec_2 ; a ) = A_1 ( a ) + A_3 ( a ) + \frac{\hat k_1 \cdot \hat k_2}{2} \frac{D_+(a)^2 }{D_+(a_i)^2}   \left( \frac{k_1}{k_2} + \frac{k_2}{k_1} \right) + ( \hat k_1 \cdot \hat k_2 )^2 \left( A_1 ( a ) -  A_3(a) \right)  \ , 
\ee
where the $A_i(a)$ are integrals over Green's functions, and $A_3 ( a)$ explicitly depends on the new non-linear source term $S( \kvec; a )$ (see \citee{Cusin:2017wjg} for details).

Now we move on to the second scheme, which we simply refer to as the EdS approximation. This is known to be a very good approximation  in $\Lambda$CDM\footnote{In $\Lambda$CDM, the EdS approximation has been shown to be accurate to better than one percent at one loop  \citees{Fosalba:1997tn,Bernardeau:1993qu,Takahashi:2008yk,Bose:2016qun}.} and for this reason it is often used in data analyses and forecasts.  
For nDGP, because the linear solutions for eqs.~\eqref{eq:Perturb1} and \eqref{eq:Perturb2} can be significantly modified by $\mu(  a)$ from their $\Lambda$CDM forms, it is not clear a priori how accurate the EdS approximation should be.  

In the EdS approximation, one assumes that the terms coming from the purely $\Lambda$CDM vertices have the EdS time dependence, while the other terms are treated by using the exact evolution equations.  Thus, one can write 
\be \label{edsapprox}
F_n(\boldsymbol{k}_1,...,\boldsymbol{k}_n ; a) =  D_+(a)^n F^{\rm EdS}_n ( \kvec_1 , \dots , \kvec_n ) + \sum_i f_i ( a ) \gamma_n^{(i)} ( \boldsymbol{k}_1,...,\boldsymbol{k}_n ) \ ,
\ee
where $F^{\rm EdS}_n$ are the usual momentum dependent kernels from $\Lambda$CDM (see for example \citee{Bernardeau:2001qr}), and the remaining term is written as a separable sum over time-dependent and momentum-dependent functions.  In particular, the functions $f_i(a)$ are the solutions of differential equations related to the linear equation of motion \eqn{lineaeq}.  

This approximation has been applied to DGP gravity and we refer the reader to App. B of \citee{Koyama:2009me} for the explicit forms of $F_2$ and $F_3$.  For example, one has
\be
\gamma_2^{(2)} ( \kvec_1 , \kvec_2) = 1 - \frac{(\kvec_1 \cdot \kvec_2)^2}{k_1^2 k_2^2}  \ ,
\ee 
and $f_2$ satisfies
\be
\left[ \frac{d^2}{d t^2} + 2 H \frac{d}{dt} - \left( 1 + \frac{1}{3 \beta} \right) \frac{3 H^2 \Omega_{ m}}{2}    \right] f_2  =  - 2 r_c^2 H^4 \left( \frac{1}{3 \beta} \right)^3 \left( \frac{ 3 \Omega_{ m}}{2} \right)^2 D_+^2 \ .
\ee

In \secref{resultssec}, we will compare the EdS approximation to using the exact Green's functions to solve perturbatively for the time dependence.

%
%
%
%

\subsection{SPT power spectra}

The power spectrum is defined through the two-point function as
\begin{equation}
\langle \delta (\bfk;a) \delta (\bfk';a)\rangle =
(2\pi)^3\delta_{\rm D}(\bfk+\bfk')\,P(k;a) \ ,
\end{equation} 
where $\langle ... \rangle$ denotes an average over initial conditions.  Assuming Gaussian initial conditions, the power spectrum can be expanded up to one loop in SPT as
\be \label{ptotal}
P(k;a) = P_{11} ( k;a) + P_{1\text{-loop}} ( k ; a ) \ .
\ee
Above, $P_{11} ( k;a) $ is the linear contribution given by 
\begin{equation}
\langle \delta_1(\bfk;a) \delta_1(\bfk';a)\rangle =
(2\pi)^3\delta_{\rm D}(\bfk+\bfk')\,P_{11}(k;a) \ ,
\end{equation} 
which, using the linear solution in \eqn{lin_sol}, can be written as 
\be
P_{11} ( k ; a ) = \left( \frac{ D_+ (a)}{D_+(a_i)} \right)^2 P^{\rm in}_k  \ , \hspace{.2in} \text{where} \hspace{.2in} (2\pi)^3\delta_{\rm D}(\bfk+\bfk')P^{\rm in}_k \equiv \langle \delta_i (\bfk) \delta_i (\bfk')\rangle  \ ,
\ee
so that $P^{\rm in}_k$ is the linear power spectrum set at some initial time.  Finally, the one-loop piece is given, in SPT, as 
\be \label{oneloop}
P_{1\text{-loop}} ( k ; a )  \equiv P_{22} ( k ; a ) + P_{13} ( k ; a ) \ , 
\ee
where 
\begin{align}
\langle \delta_{2}(\bfk;a) \delta_2(\bfk';a)\rangle &=
(2\pi)^3\delta_{\rm D}(\bfk+\bfk')\,P_{22}(k;a) \ , \label{eq:psconstraint0} \\
2 \langle \delta_1(\bfk;a) \delta_3(\bfk';a) \rangle &=
(2\pi)^3\delta_{\rm D}(\bfk+\bfk')\,P_{13}(k;a) \ .
\label{eq:psconstraint1}
\end{align}
In \sect{eftsec} we will discuss the counterterms which, in the EFTofLSS, appear in \eqn{oneloop}.

 Now let us move on to the explicit expressions of the one-loop contribution in perturbation theory.  First we will discuss the case where the time dependence is solved using the exact Green's functions, and then we will discuss the EdS approximation.  As shown in \citee{Cusin:2017wjg}, the solution using the exact Green's functions is 
\begin{align} \label{p22foryou}
P_{22} ( k; a) & = \int \momspmeas{\bfq} \int_0^{a}  d a_2 \int_0^{a_2} d a_1 \, \,  p_{22}(a, a_1 , a_2 ; \kvec , \qvec) \ , \\ 
 P_{13} ( k; a) & = \int \momspmeas{\bfq} \int_0^{a}  d a_2  \left(  p_{13}^{(1)} (a , a_2 ; \kvec , \qvec)  +   \int_0^{a_2} d a_1 \, \,  p_{13}^{(2)}(a, a_1 , a_2 ; \kvec , \qvec)  \right)  \ , \label{p13foryou}  
\end{align}
where the integrands in these expressions are given by
\begin{align} \label{p22text2}
p_{22}(a, a_1 , a_2 ; \kvec , \qvec) & \equiv  \sum_{i = 1}^7 T^{(22)}_i (a, a_1 , a_2) F^{(22)}_i ( \kvec , \qvec) \ ,  \\
p_{13}^{(2)}(a, a_1 , a_2 ; \kvec , \qvec) & \equiv  \sum_{i = 1}^{10} T^{(13)}_i (a, a_1 , a_2) F^{(13)}_i ( \kvec , \qvec) \; ,\\
p_{13}^{(1)}(a , a_2 ; \kvec , \qvec) & \equiv  T^{(13)}_{11} (a , a_2) F^{(13)}_{11} ( \kvec , \qvec)  \;.   \label{p22text3}
\end{align}
Although we will skip most of the details and refer the reader to \citee{Cusin:2017wjg} and \appref{exacttimeapp} for specific expressions, let us make a few quick comments about \eqnstwo{p22foryou}{p13foryou}.   First, notice that, in addition to the normal integral over the internal momentum $\bfq$, there are also integrals over time in \eqnstwo{p22foryou}{p13foryou} due to the fact that we are using the exact Green's functions.  Additionally, each individual term in \eqn{p22text2} - \eqn{p22text3} is written as the product of a time-dependent function $T^{(\sigma)}_i$ and a momentum dependent function $F^{(\sigma)}_i$; this is possible only when $\mu(a)$ does not depend on $k$, and the linear equations are scale independent.  The momentum dependent kernels $F^{(\sigma)}_i $ are given by a mixing of the standard $\alpha( \bfk_1 , \bfk_2)$ and $\beta( \bfk_1 , \bfk_2)$ kernels from $\Lambda$CDM and the new kernel $\gamma_2(\bfk_1 , \bfk_2)$, and are proportional to two factors of the initial power spectrum $P^{\rm in}$.  The time dependent functions $T^{(\sigma)}_i$ depend on the Green's functions and the nDGP function $\beta(a)$.  In particular, in the expression for $p_{22}$, the terms in the sum for $i = 1,\dots,4$ are the normal $\Lambda$CDM terms, and the terms for $i = 5,6,7$, are the non-linear terms that appear in DGP.  Similarly, in the expression for $p_{13}$, the terms with $ i = 1,\dots,6$ are the normal $\Lambda$CDM terms, and the terms with $i=7,\dots,11$ are the non-linear terms from DGP.\footnote{To give some intuition, let us consider an example. For the terms that are also present in $\Lambda$CDM, we have 
\begin{align}
F^{(13)}_1 ( \kvec , \qvec)  = 4 \,  \alpha_s ( \kvec , \qvec) \, \alpha ( - \qvec , \kvec + \qvec) \, P^{\rm in}_{k}\, P^{\rm in}_{q} \ , \hspace{.1in} \text{and} \hspace{.1in} 
T^{(13)}_1 ( a , a_1 , a_2)  = K( a , a_1 , a_2 )\,   G^\delta_1 ( a , a_2)  G^\delta_1 ( a_2 , a_1)  \ , 
\end{align}
and for the terms that are present in nDGP because of screening, we have
\begin{align}
F^{(13)}_8 ( \kvec , \qvec)  & = 4 \,  \alpha_s ( \kvec , \qvec) \,  \gamma_2 (  \kvec + \qvec , - \qvec ) \, P^{\rm in}_{k}\, P^{\rm in}_{q} \ ,  \\ 
T^{(13)}_8 ( a , a_1 , a_2)  &=  2 \,  \left( \frac{3 \, \Omega_m (a_2) }{2}  \right) \mu_{\Phi,2}( a_2) K( a , a_1 , a_2 )  \, f_+(a_2)^{-1}\,   G^\delta_2 ( a , a_2)  G^\delta_1 ( a_2 , a_1)   \ , 
\end{align}
where $\alpha_s ( \bfk , \bfq ) = \frac{1}{2} ( \alpha ( \bfk , \bfq ) + \alpha(\bfq , \bfk))$, and 
\be
K( a , a_1 , a_2 ) = \frac{ a_1 a_2 D_+(a ) D_+ ( a_1) D'_+(a_1) D'_+(a_2)}{D_+(a_i)^4} \ .
\ee

}

%
%
%

\subsection{Effective field theory of large-scale structure} \label{eftsec}

From the viewpoint of the EFTofLSS, the matter evolution equations in \eqn{eq:Perturb1} and \eqn{eq:Perturb2} are incomplete \citees{Baumann:2010tm,Carrasco:2012cv}.  This is because the description of dark matter as a perfect fluid is only correct on the largest scales.  On smaller scales, the evolution of the long-wavelength modes of interest can be significantly affected, through the non-linear couplings, by short-wavelength (i.e. UV) modes.  Because the short-wavelength modes are not under perturbative control, it is not possible to predict exactly what these effects are.  In the context of $\Lambda$CDM, it was shown in \citee{Baumann:2010tm} that these effects, however, enter as the divergence of an effective stress tensor in \eqn{eq:Perturb2}.\footnote{In \cite{Cusin:2017mzw} it was also explicitly shown that this is the case in the quasi-static limit for Horndeski-type theories in the EFTofDE.  We expect the same to be true in the quasi-static limit of nDGP that we consider in this paper.  Indeed, in this limit, the scalar field is non-dynamical so that there can be, for example, no relative velocity effects.}  Then, using the equivalence principle, one can write down the most general form of the effective stress tensor as a controlled expansion in terms of powers and derivatives of the long-wavelength fields; this procedure fixes the possible dependence in $k$-space of the UV effects.  The specific details about the UV physics is encoded in a set of time-dependent couplings which are undetermined by the theory and must be fit either by simulations or observations.

For the one-loop computation, this procedure introduces one extra term to the right-hand side of \eqn{eq:Perturb2}, which we write as\footnote{We ignore stochastic contributions, which are generally negligible in a one-loop computation.}
\be \label{cteom}
+ 9 \, ( 2 \pi ) c_{\delta,1}^2 ( a ) \frac{k^2}{\knl^2} \delta( \bfk; a) \ .
\ee
Notice that we have written the free dimensionful coefficient as a dimensionless function of time  $c_{\delta ,1}^2(a)$, called the speed of sound, and a dimensionful parameter $\knl^2$, called the non-linear scale.  In this way, if $c_{\delta,1}^2$ is of order unity, then $\knl$ is the scale that suppresses derivatives in the theory: it is the strong-coupling scale of the EFT.   

The new term \eqn{cteom} in the equations of motion can be treated as a (third-order) source term, which, using the Green's functions, gives a contribution to the expansion of $\delta$ as  
\be
\delta^{\rm ct} ( \kvec ; a ) = - (2 \pi) c_{s}^2 ( a ) \frac{k^2}{\knl^2} \delta_1 ( \kvec ; a )  \ ,
\ee
where we have introduced the speed of sound parameter $c_{s}^2$, defined as 
\be
c_{s}^2 ( a ) \equiv \int^a d a ' \, G_2^\delta ( a , a') 9 c_{\delta , 1}^2 ( a') \frac{D_+(a')}{D_+(a)} \ .
\ee
 This is the quantity that enters in the power spectrum. Indeed,  the expression for the one-loop power spectrum in the EFTofLSS is 
\begin{equation}
P^{{\rm EFT}} ( k;a) = P_{11} ( k ; a ) + P_{1\text{-loop}}( k;a)  + P^{\rm ct}_{13}(k;a) \ ,
\label{eftps}
\end{equation}
where the last term is given by
\be
\label{P13}
2 \langle \delta_1(\bfk;a) \delta^{\rm ct} (\bfk';a) \rangle = (2\pi)^3\delta_{\rm D}(\bfk+\bfk')\,P^{\rm ct}_{13}(k;a) \;, \qquad P^{\rm ct}_{13}(k;a) = - 2 (2\pi) c_s^2(a)  \frac{k^2}{k_{\rm NL}^2}P_{11}(k;a) \;.
\ee
Because we will be working directly with the power spectrum in this paper, we will use $c_s^2(a)$ (instead of $c_{\delta , 1}^2 (a)$) as the free parameter.

%
%
%

\subsection{Resummation schemes} \label{resummationsec}

The EFTofLSS puts us in a position to probe smaller, information rich scales in upcoming LSS surveys.  Although the UV reach of the theory was improved by using the EFTofLSS, it was noted early that the computations showed an oscillatory residual, with an amplitude of approximately $2\%$, when compared to N-body results \citees{Carrasco:2012cv,Carrasco:2013mua}.  These residuals are due to the insufficient treatment of the BAO oscillations by Eulerian perturbation theory, and various resummation methods tackling this issue have been proposed \citees{Matsubara:2007wj,Senatore:2014via,Baldauf:2015xfa,Vlah:2015sea,Vlah:2015zda}.  With the onset of large volume, stage IV surveys, highly accurate theoretical templates are required in order to achieve unbiased constraints on cosmology and gravity, and so the choice of resummation method may introduce systematic biases in parameter constraints. To address this, we compare two such methods in this work.

\subsubsection{R1 resummation}
First, we consider a systematic method to resum the long-wavelength displacement modes \citee{Senatore:2014via} (see also \citees{Lewandowski:2015ziq,Senatore:2017pbn} for further developments and applications), which we will refer to as method ``R1''.  In this approach, one recognizes three important expansion parameters in Eulerian perturbation theory, which appear in the loop expansion
\begin{align} \label{epsilondef}
 \epsilon_{s < } ( k )  \equiv  k^2 \int_0^k \frac{d^3 k'}{( 2 \pi)^3} \frac{ P_{11} ( k ' )}{k'^2}   \ , \hspace{.1in}   \epsilon_{\delta < }  ( k ) \equiv  \int_0^k \frac{d^3 k'}{( 2 \pi)^3} P_{11} ( k ' )  \ , \hspace{.1in} \text{and} \hspace{.1in}  \epsilon_{s >} ( k ) \equiv  k^2 \int_k^\infty \frac{d^3 k'}{( 2 \pi)^3} \frac{ P_{11} ( k ' )}{k'^2}  \ .
\end{align}
As one can see in \eqn{epsilondef}, $\epsilon_{s <}$ is related to IR displacements, $\epsilon_{ \delta < }$ is related to IR density fluctuations and $\epsilon_{s >}$ is related to UV displacements.  Eulerian perturbation theory expands equally in all three of these parameters, but the BAO oscillations make $\epsilon_{s < }$ large on the scales of interest, so the IR-resummation is a method to treat non-perturbatively the modes related to IR displacements.  


Although we present many more details in \appref{irresumapp}, here we simply provide the final form of the resummation and make a few comments.  Ultimately, the IR-resummation, expanded to one-loop order in $\epsilon_{ \delta < }$ and $\epsilon_{s >}$ but kept to all orders in $\epsilon_{s <}$, is given by
\begin{align} \label{r11}
P ( k; a ) \Big|_1 & =  \int \frac{d k' \, k'^2}{2 \pi^2} \left(  M_{||_{1}} ( k , k' ; a ) \, P_{11} ( k' ;a ) +  M_{||_{0}} ( k , k' ; a ) \, P_{1\text{-loop}} ( k' ;a )    \right)  \ ,
\end{align}
where $P_{11}$ and $P_{1\text{-loop}}$ are the Eulerian power spectra, including EFT counterterms, and the $ M_{||_{i}} ( k , k' ; a )$ kernels contain the information about the IR modes that are being resummed, see \eqn{sbt} for the explicit forms.  In particular, in \eqn{r11}, the linear IR modes are resummed exactly.\footnote{A recent work has shown that resumming the first non-Gaussian IR modes makes only a very small improvement in $\Lambda$CDM \citee{Senatore:2017pbn}, so we are justified in keeping only the linear displacements.}  The challenge to using \eqn{r11} is computing $ M_{||_{i}} ( k , k' ; a )$ quickly without making uncontrollable approximations (for one method, particularly relevant in redshift space, see \citee{Lewandowski:2015ziq}).  For the rest of this paper, we will write
\be\label{r1}
P_{\rm R1} ( k;a) \equiv P ( k; a ) \Big|_1 \ ,
\ee
where the subscript $``\rm R1"$ simply means that this is the first resummation method that we will consider in this paper.

\subsubsection{R2 resummation}
The second resummation scheme that we consider, which we call ``R2,'' was proposed by \citee{Baldauf:2015xfa} (see also \citees{Vlah:2015sea,Vlah:2015zda,Blas:2016sfa,delaBella:2017qjy} for further developments and applications). This approach splits the linear power spectrum into `wiggle' and `no-wiggle' parts, where the wiggle component contains oscillations associated with the BAO peak, and the no-wiggle component contains the broadband power, i.e.  
\begin{equation} \label{wignowig}
P_{11} ( k ; a ) = P_{11}^{\rm w} ( k ; a ) + P_{11}^{\rm nw}(k;a) \ . 
\end{equation}
This splitting is somewhat arbitrary, so here we follow \citee{delaBella:2017qjy} and choose 
\begin{equation}
P^{\rm nw}_{11}(k;a) = \frac{P_{\rm EH}(k;a)}{\sqrt{2\pi \lambda^2}} \int \frac{dq}{q} \frac{P_{11}(q;a)}{P_{\rm EH}(q;a)} {\rm exp} \left( -\frac{(\ln{(k/q)})^2}{2\lambda^2}\right),
\end{equation}
where $P_{\rm EH}$ is the Eisenstein and Hu no-wiggle spectrum \citee{Eisenstein:1997jh} and $\lambda$ is a dimensionless parameter that sets the size of the filter window, taken to be $\lambda = 0.25 ( k / k_{\rm pivot})^{0.04}$, where $k_{\rm pivot } = 0.05 \unitsk$ as in \citee{delaBella:2017qjy}.\footnote{The parameter $\lambda$ was used in \citee{Vlah:2015zda} where it is explained in more detail.  The explicit expression used in this work was taken from \citee{delaBella:2017qjy}.  Since $\lambda$ sets the size of the window which is filtering over the BAO oscillations, it controls how much of the real BAO oscillations are present in the wiggle power spectrum, which gets resummed, and how much are present in the no-wiggle power spectrum, which is not resummed (i.e. one would have $P^{\rm nw}_{11} (k)\rightarrow P_{11} ( k ) $ as $\lambda \rightarrow 0$).  This is a part of the ambiguity related to defining the wiggle and no-wiggle parts of the power spectrum.}  For the one-loop terms, $P_{1\text{-loop}}^{\rm nw}$ is defined to be the standard Eulerian expression, but with $P_{11}$ replaced with $P_{11}^{\rm nw}$.  The wiggly part of the one-loop piece can then be defined as $P_{1\text{-loop}}^{\rm w} \equiv P_{1\text{-loop}} - P_{1\text{-loop}}^{\rm nw}$.  Then, with these definitions  the expression for the resummed power spectrum in this scheme at one loop is 
\begin{equation}
P_{\rm R2} (k;a) =  P_{\rm nw}^{\leq 1}(k;a) + {\rm exp} \left( -\frac{k^2}{2} \langle \langle  A_0^{\rm nw} \rangle \rangle \right) \left[ P_{\rm w}^{\leq 1}(k;a)  + \frac{k^2}{2}\langle \langle  A_0^{\rm nw} \rangle\rangle P^{\rm w}_{\rm 11} ( k ; a)\right]  \ , 
\label{r2}
\end{equation}
where $P^{\leq 1} \equiv P_{11} + P_{1\text{-loop}} + P^{\rm ct}_{13}$ denotes the Eulerian power spectrum up to and including one loop, and $\langle \langle  A_0^{\rm nw} \rangle\rangle$ is an average of the two-point correlation of the linear displacement field over the range of scales where the BAO oscillations are supported. This can be approximated by 
\begin{equation}
 \langle \langle  A_0^{\rm nw} \rangle\rangle = \frac{1}{\pi^2} \frac{1}{q_{\rm max}^2 - q^2_{\rm min}} \int^{q_{\rm max}}_{q_{\rm min}} dq \, q^2 \int dk \,  P_{11}^{\rm nw}(k;a)[1-j_0(kq)] \ ,
\end{equation}
where $j_0$ is the 0th-order spherical Bessel function. $q_{\rm max}$ and $q_{\rm min}$ are chosen to encompass all scales over which the oscillations are supported. We choose the same limits as \citee{delaBella:2017qjy}, $q_{\rm max} = 300\mbox{Mpc}\, h^{-1}$ and $q_{\rm min} = 10 \mbox{Mpc} \, h^{-1}$. 

The relation between resummation schemes R1 (\eqn{r11}) and R2 (\eqn{r2}) was briefly discussed in \citee{Vlah:2015sea}.   Here, however, we would like to make one conceptual point regarding the difference between the schemes R1 and R2.  As discussed in \citee{Senatore:2014via}, the scheme R1 exactly resums the linear displacements; the error in the computation is that there are indeed displacements due to short modes.   However, because this error is due to short modes, it will be recovered order by order in perturbation theory.  On the other hand, the splitting of the power spectrum into wiggle and no-wiggle parts in \eqn{wignowig} introduces a systematic error which is not recovered in perturbation theory as one goes to higher loop order.  Although we will show next that these two methods are quite comparable numerically, we consider the resummation scheme R1 to be the most theoretically justified for the above reason.  Direct comparisons of these schemes is made in the following section.

\section{Results} \label{resultssec}
In this section we compare the predictions of various theoretical approaches with power spectrum measurements from  MG-PICOLA \cite{Winther:2017jof} simulations in the normal branch of DGP gravity and GR. MG-PICOLA is based on a parallel COLA  implementation (PICOLA) (see \citee{Howlett:2015hfa} for details). MG-PICOLA is relatively computationally inexpensive, but this advantage comes at the price of limited accuracy (when compared to a same resolution N-body run). This limited accuracy has been well quantified and has been shown to be sub percent up to $k\sim 1 \unitsk$ for the models considered \cite{Winther:2017jof} and leads to a steady suppression of power at small scales.  Similar results were found for another COLA implementation dealing with symmetron and chameleon screened models \cite{Valogiannis:2016ane}.

The background cosmology is taken from WMAP9 \citee{Hinshaw:2012aka}: $\Omega_{m0} = 0.281$, $h=0.697$, $\Delta \zeta^2 = 2.538 \times 10^{-9}$, and $n_s=0.971$.  The box  length of the simulations is $1024 \mbox{Mpc} \, h^{-1}$ with $1024^3$ dark matter particles and a starting redshift of $z=19$.  Our simulations were run with $N = 1024$ grid-cells in each dimension corresponding to a Nyquist frequency of $k = 1.57 \unitsk$.  Because the grid provides a finite UV cutoff, we do not need to employ explicit force softening.  By comparing to high resolution $N$-body simulations we have found that the power spectrum of our COLA simulations is accurate to better than $1\%$ (at $z = 0.0$) for modes $k < 0.6 \unitsk$. The initial conditions were generated using second order Lagrangian perturbation theory, and both nDGP and $\Lambda$CDM simulations begin with the same initial seeds. The theoretical predictions are compared to the average of 20 such simulations. For the nDGP simulations we make use of various values of the nDGP parameter $\Omega_{\rm rc}$ but focus on the $\Omega_{\rm rc} = 0.438$ case as it offers comparably significant deviations to the $\Lambda$CDM case. Comparisons are made at the redshifts of $z=0,0.5$, and $1$, and loop integrals in the perturbative expansion are performed with an IR cutoff of $k_{\rm IR} = 10^{-4} \unitsk$ and a UV cutoff of $k_{\rm UV} = 10 \unitsk$.  We begin by looking at two common approximations.

%
%
%

\subsection{Time dependence and screening approximations} 
\label{sub1}

Before looking at simulation data, we first consider the effects of two common approximations used in LSS survey analyses, in the framework of SPT ($\bar c_s^2 =0$).  First, regarding the time dependence, we look at the difference between using the exact Green's functions and using the EdS approximation described in \secref{linearsolsec}.  Then we look at the effect of ignoring the screening terms in the equations of motion, i.e. of setting $S(\kvec ; a ) = 0$ in \eqn{eq:poisson1}; we call this approximation the unscreened approximation, or UsA.  We start with the time dependence.

 \begin{figure}[h]
  \captionsetup[subfigure]{labelformat=empty}
  \centering
   \subfloat[]{\includegraphics[width=17cm]{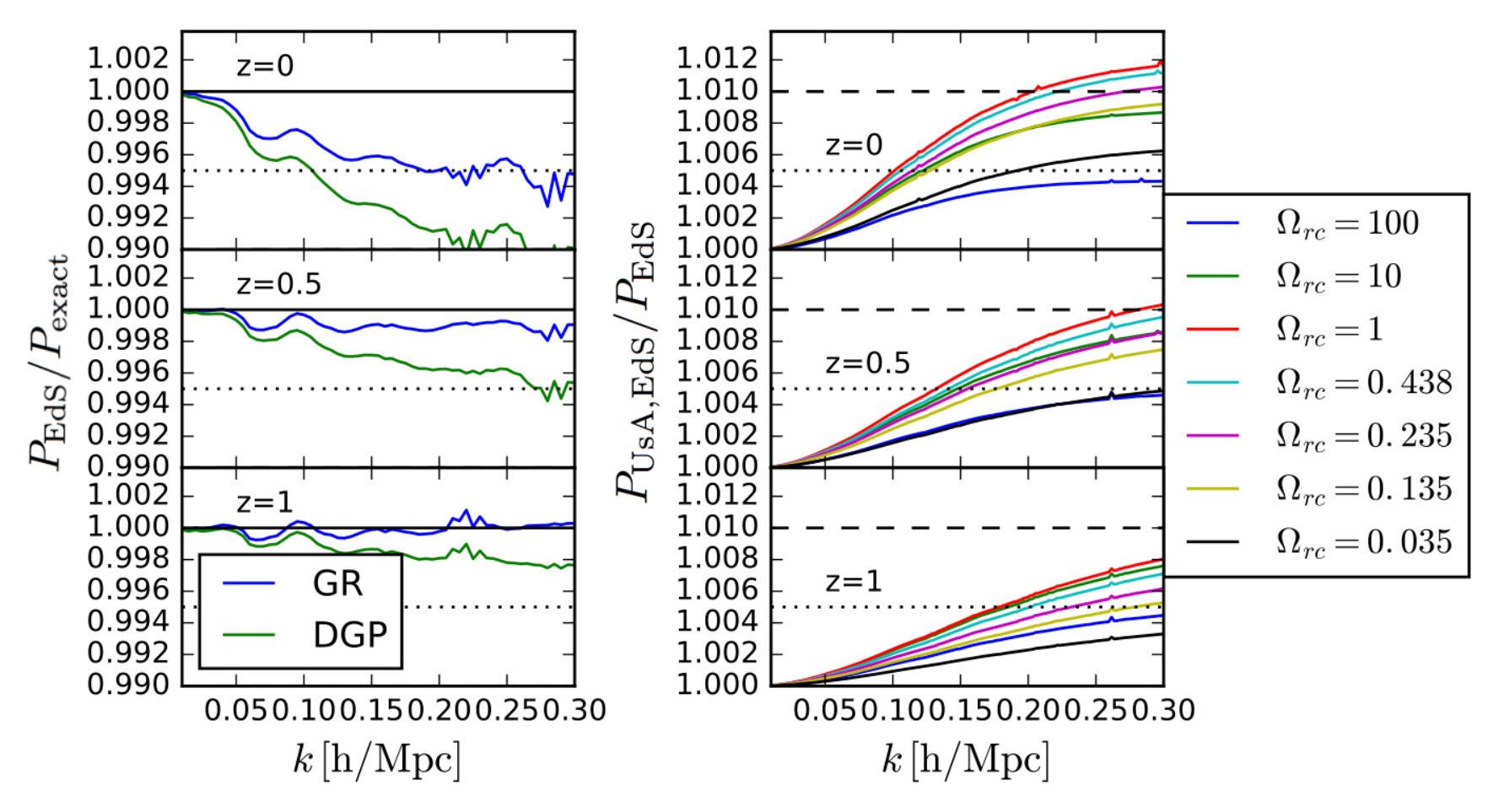}} \quad
  \caption[CONVERGENCE ]{\footnotesize {\bf Left}: The ratio of the EdS to the exact time dependence one-loop power spectrum for $\Lambda$CDM (blue) and DGP (green) with $\Omega_{\rm rc}=0.438$, in SPT.  We see that the EdS approximation is valid to less than a percent at $z=0$ and $k \approx 0.2 \unitsk$. {\bf Right}: The ratio of the UsA+EdS to the EdS one-loop power spectrum for varying values of $\Omega_{ \rm rc}$. We show results for $z=0,0.5$, and $1$.}
\label{usaandeds}
\end{figure}

 \begin{figure}[h]
  \captionsetup[subfigure]{labelformat=empty}
  \centering
  \subfloat[]{\includegraphics[width=17cm, height=8.1cm]{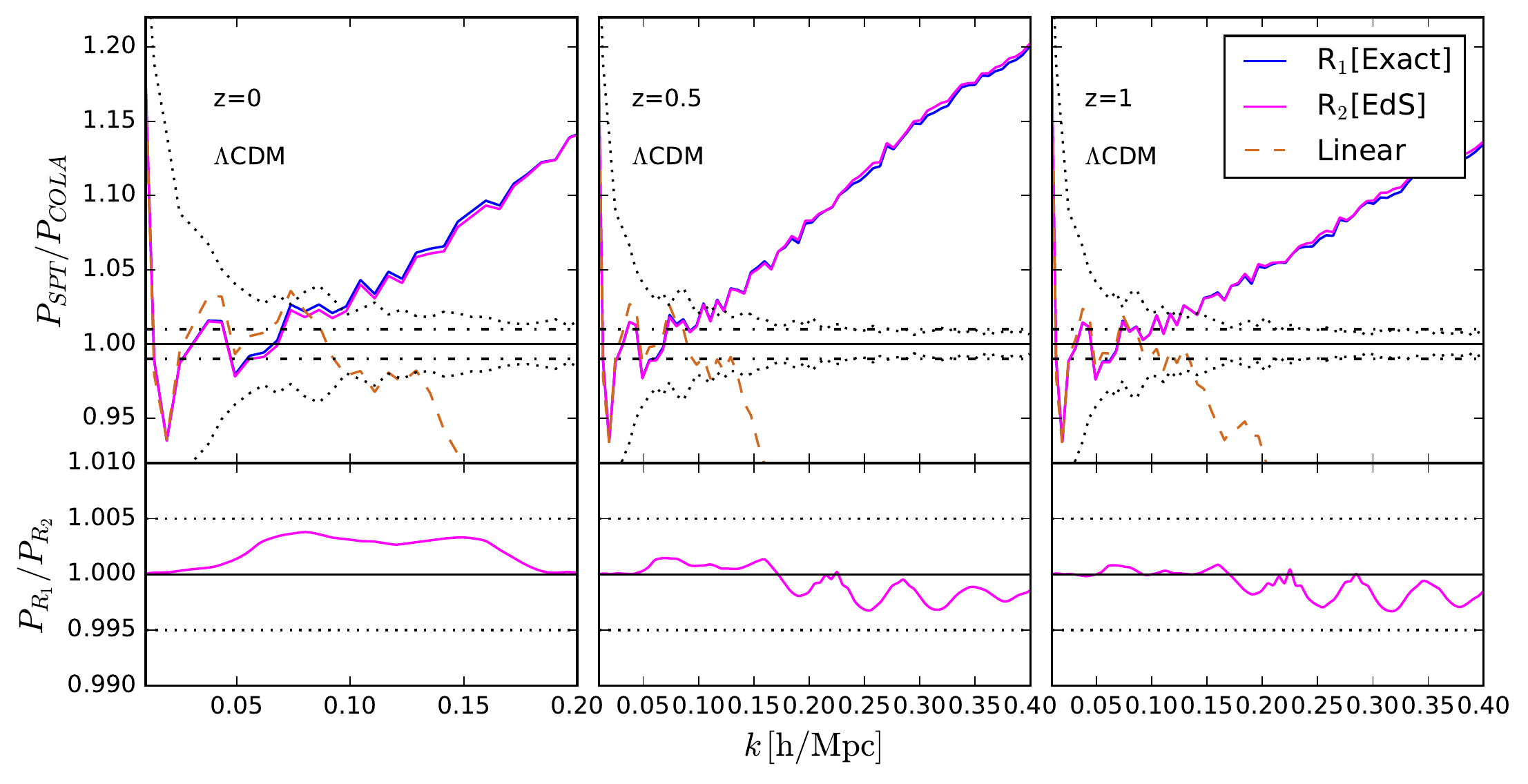}} 
  \caption[CONVERGENCE ]{ \footnotesize $\Lambda$CDM comparisons.  {\bf Top}: The ratio of the resummed one-loop matter power spectrum computed using eq.~(\ref{r1}) (blue) and eq.~(\ref{r2}) with EdS (magenta) for $\Lambda$CDM, in SPT ($ \bar c_s^2 = 0$), to the COLA data.  The linear prediction is the dashed orange line.  The black dotted line is the variance of the 20 COLA realizations. {\bf Bottom}: The ratio of the blue curve to the magenta curve from the top panels. We see that the difference is less that $0.5\%$ for the different resummation schemes.  This is shown for $z=0$ (left), $z=0.5$ (center), and $z=1$ (right).}
\label{grspt}
\end{figure}

The left-hand side of Fig.~\ref{usaandeds} shows the ratio of the EdS to the exact computation, both described in \secref{linearsolsec}.  As has been shown in the literature (see for example \citees{Fosalba:1997tn,Bernardeau:1993qu,Takahashi:2008yk,Bose:2016qun}), in a $\Lambda$CDM cosmology the EdS approximation is valid to less than one percent on the typical scales of interest. From \figref{usaandeds}, we see that the EdS approximation is slightly worse for nDGP, but still sub percent, and works to systematically suppress power.

The right-hand side of Fig.~\ref{usaandeds} quantifies the UsA approximation for varying values of the nDGP parameter $\Omega_{\rm rc}$. It shows the ratio of the combined UsA and EdS approximations to the EdS-only approximation at one loop. We see that neglecting the screening terms gives an effect with enhanced power at smaller scales, which is opposite of the effect of the EdS approximation shown in the left panel.  The effect of the UsA should be typical of most modified gravity theories as these screening terms aim to suppress deviations from GR as we approach the screened regime.  We see that this contribution saturates for $\Omega_{\rm rc}=1$, where it is a $1\%$ effect at $k=0.2 \unitsk$ at $z=0$.  Indeed, we did not expect the nDGP non-linear vertices to contribute significantly because the non-linear coupling $\mu_{2}$ (see  \eqn{gamma2}) is so small.  To see this, consider the largest modification, which is for $\Omega_{\rm rc} = 1$.  Then, from \eqn{betadef} we have that $1 / ( 3 \beta ) \approx 1/6$ and $H_0^2 r_c^2 = 1/4$.  Then, using \eqn{gamma2}, we have that $\mu_{2} = -2 H^2 r_c^2 ( 3 \beta)^{-3}$, which has a current-day value of $\mu_{2} ( a_0 ) \approx  - 1/(2 \cdot 6^3) = -1/432$.  Thus, the main difference on non-linear scales is due to the modified growth factor $D_+$ in the normal $\Lambda$CDM non-linear terms.  Next, we move on to compare directly with the COLA simulations.

%
%
%

\subsection{One-loop SPT comparisons} \label{comparisonsec1}
In this subsection, we compare the various matter power spectrum predictions with non-linear simulations.  We will investigate the effects of the UsA and EdS approximations, as well as the two different resummation schemes R1 and R2.  Throughout this subsection, when using the R1 resummation method, \eqn{r1}, we will use the exact time dependence for the loops, and when using the R2 resummation method, \eqn{r2}, we will use the EdS approximation for the loops.  We do this because using the exact time dependence and R1 resummation scheme represents the most theoretically justified computation, but is computationally heavier.  On the other hand, using the EdS approximation with resummation R1 is the fastest computational strategy, so we are most interested in seeing how these two scenarios compare.

In \figrefs{grspt}{dgpspt}, we look at the SPT power spectra.  \figref{grspt} shows the comparisons for $\Lambda$CDM. The top panels show the ratio of the COLA measurement to the predictions of eq.~(\ref{r1}) (blue) and eq.~(\ref{r2}) (cyan) with $ c_s^2=0$ (SPT). The dotted black curves show the variance over the 20 COLA realizations. The dashed orange curve shows linear theory and the dot-dashed lines delimit the $1\%$ accuracy region, this being of particular relevance for upcoming surveys.  At all redshifts, we find that the resummation schemes are consistent within to $0.5\%$.

 \begin{figure}[h]
  \captionsetup[subfigure]{labelformat=empty}
  \centering
  \subfloat[]{\includegraphics[width=17cm, height=8.1cm]{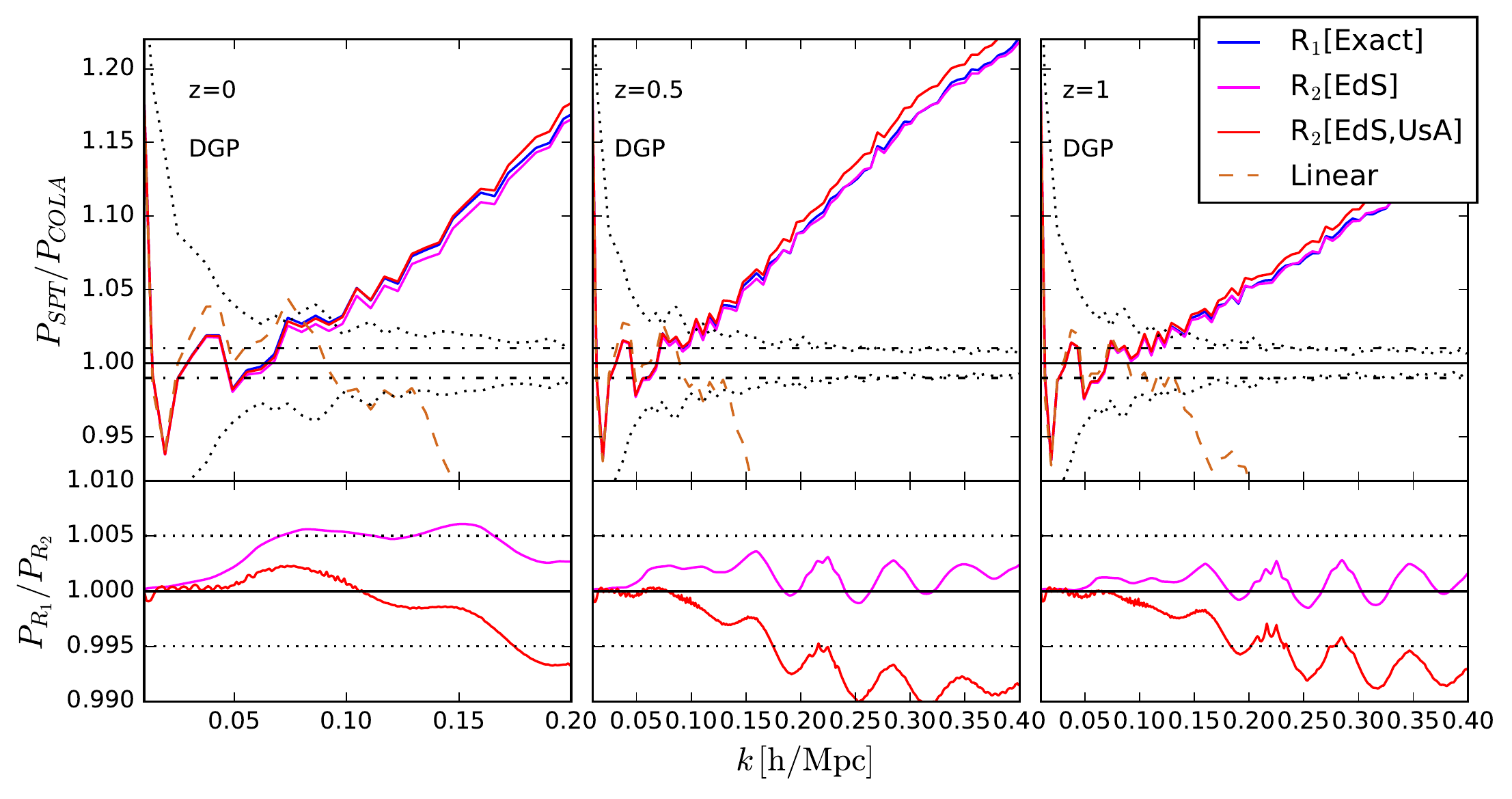}} 
  \caption[CONVERGENCE ]{ \footnotesize nDGP comparisons. {\bf Top}: The ratio of the resummed one-loop matter power spectrum computed using eq.~(\ref{r1}) (blue), eq.~(\ref{r2}) with EdS (magenta) and eq.~(\ref{r2}) with EdS and UsA (red) for nDGP in SPT ($\bar{c}_s^2=0$), to the COLA data. The linear prediction is shown as the dashed orange curve.  The black dotted line is the variance of the 20 COLA realizations. {\bf Bottom}: The ratio of the blue curve to the magenta curve (magenta) and the ratio of the blue curve to the red curve (red) from the top panels. We see that the maximum difference on the scales of interest is less than $1\%$.  This is shown for $z=0$ (left), $z=0.5$ (center), and $z=1$ (right).}
\label{dgpspt}
\end{figure}
\figref{dgpspt} shows the comparisons for nDGP. We have included an additional red curve which includes the addition of the UsA approximation for resummation method R2.  The EdS and UsA approximations as well as different resummation methods still do not amount to more than a $1\%$ difference when compared to the exact computation.

%
%

\subsection{One-loop EFT comparisons} \label{comparisonsec2}

Moving to the EFTofLSS framework, we now employ eq.~(\ref{eftps}) and fit $ c_s^2$ (see eq.~\eqref{P13}) to the COLA data. 
 When reporting numerical values, we will typically use the dimensionful parameter 
\be
\bar c_s^2 ( a ) \equiv  c_s^2 (a ) / \knl^2 \;.  
\ee
We fit by following the scheme described in \citee{Foreman:2015lca}, details of which are included in \appref{fittingproc}.  To discuss the fits, it is useful to introduce two scales, $\kfit$ and $\kreach$.  The scale $\kfit$ is the maximum $k$ that is included in the least $\chi^2$ fit to determine the best fit value of $\bar c_s^2$, and $\kreach$ is the scale at which, given this value of $\bar c_s^2$, the prediction fails with respect to the non-linear data.  This last scale necessarily depends both on the error bars on the data, and on the theoretical errors, as we will discuss below.

\begin{figure}[h]
  \captionsetup[subfigure]{labelformat=empty}
  \centering
  \subfloat[]{\includegraphics[width=17cm, height=8.1cm]{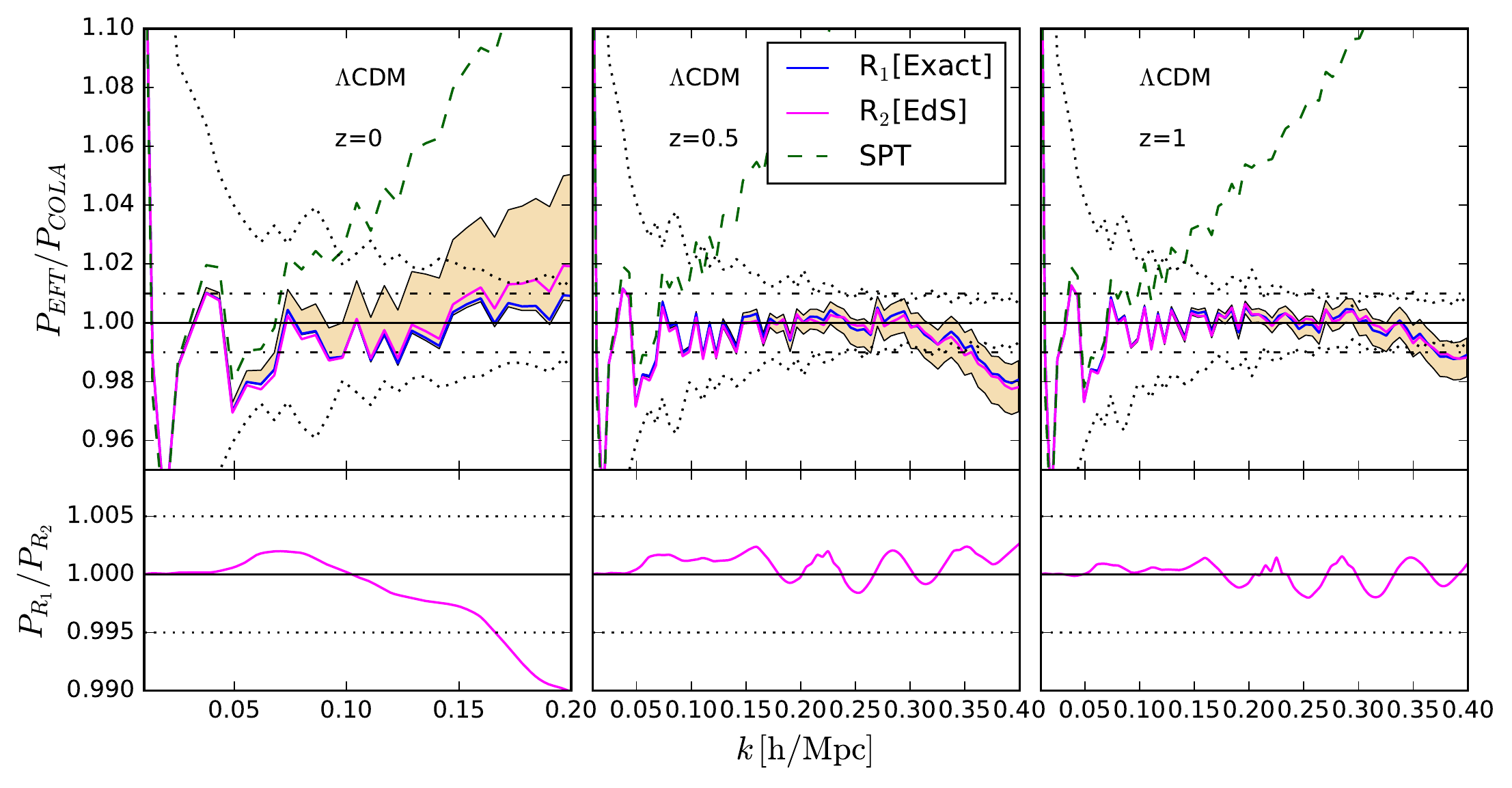}} 
  \caption[CONVERGENCE ]{ \footnotesize EFT $\Lambda$CDM comparisons.   {\bf Top}:  The EFT resummed one-loop matter power spectrum computed using eq.~(\ref{r1}) (blue) and eq.~(\ref{r2}) with EdS (magenta) for $\Lambda$CDM (each with their own best fit $\bar c_s^2$), divided by the COLA data. The one-loop SPT prediction is the dashed green line, and the yellow band indicates the theoretical error, as described in \appref{fittingproc}.  The black dotted line is the variance of the 20 COLA realizations. {\bf Bottom}: The ratio of the blue curve to the magenta curve from the top panels. We see that the difference is less that $1\%$ between the two resummation schemes.  This is shown for $z=0$ (left), $z=0.5$ (center) and $z=1$ (right).  The values for $ \bar{c}_s^2$ with their errors are shown in Tab.~\ref{cs2list}. }
\label{greft}
\end{figure}

\begin{figure}[h]
  \captionsetup[subfigure]{labelformat=empty}
  \centering
  \subfloat[]{\includegraphics[width=17cm, height=8.1cm]{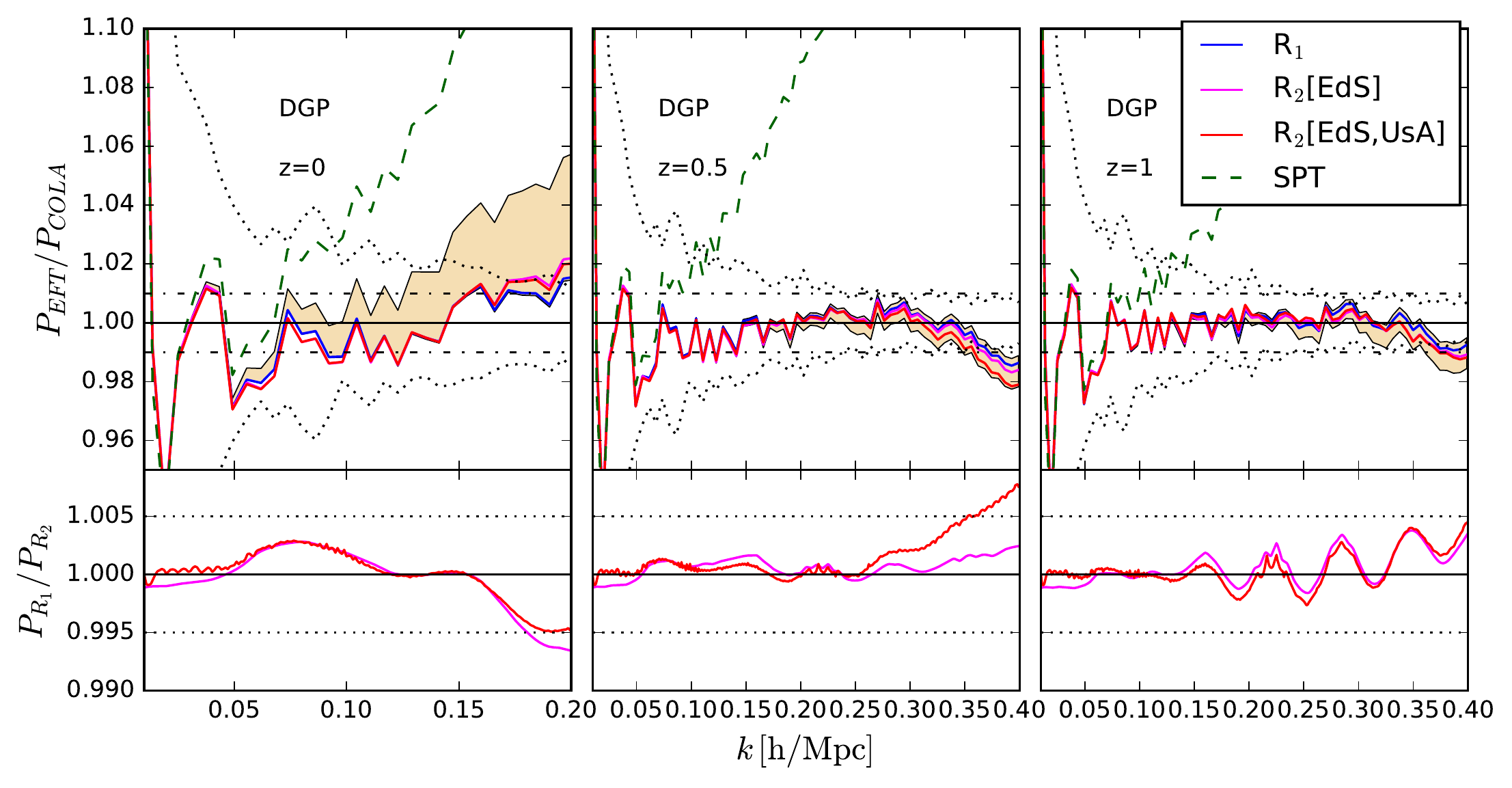}}
  \caption[CONVERGENCE ]{\footnotesize EFT nDGP comparisons.  {\bf Top}:  The EFT resummed one-loop matter power spectrum computed using eq.~(\ref{r1}) (blue), eq.~(\ref{r2}) with EdS (magenta), and eq.~(\ref{r2}) with EdS and UsA (red)  for nDGP (each with their own best fit $\bar c_s^2$), divided by the COLA data. The one-loop SPT prediction is the dashed green line, and the yellow band indicates the theoretical error, as described in \appref{fittingproc}.  The black dotted line is the variance of the 20 COLA realizations. {\bf Bottom}: The ratio of the blue curve to the magenta curve (magenta), and the ratio of the blue curve to the red curve (red) from the top panels. We see that the difference is less that $1\%$ between the two resummation schemes.  This is shown for $z=0$ (left), $z=0.5$ (center) and $z=1$ (right).  The values for $ \bar {c}_s^2$ with their errors are shown in Tab.~\ref{cs2list}. }
\label{dgpeft}
\end{figure}

The results for $\Lambda$CDM and nDGP are shown in \figref{greft} and \figref{dgpeft} respectively.  In these plots, the yellow band is the estimated theoretical error, which is described in detail in \appref{fittingproc}.  Briefly, the theoretical error is determined by the $1\sigma$ variance in the $\chi^2$ fitted value of $\bar c_s^2$ at $0.75 \kfit$.  In \appref{fittingproc}, we explain why, given the data that we have, this method is preferred over, for example, estimating the two-loop contribution.  Given this theoretical error, the scale $\kreach$ is given by when the yellow band goes outside of the error bars on the data, so that one is no longer certain that the prediction lies within the data errors.

For $\Lambda$CDM (\figref{greft}) we find that $k_{\rm reach} ( z = 0 ) \approx 0.14\unitsk$, $\kreach ( z = 0.5 ) \approx 0.30\unitsk$, and $\kreach(z=1) \approx 0.34 \unitsk$.  On the other hand, for nDGP (\figref{dgpeft}), we find $k_{\rm reach} ( z = 0 ) \approx 0.14\unitsk$, $\kreach ( z = 0.5 ) \approx 0.35 \unitsk$, and $\kreach(z=1) \approx 0.36 \unitsk$.  We show all values of $\bar{c}_s^2$ for all curves in Tab.~\ref{cs2list} with their $1\sigma$ errors.  From Tab.~\ref{cs2list} we see that the values of $\bar c_s^2$ are consistent over the various approximations, resummation schemes, and models.  Overall, the resummation schemes are consistent to within $1\%$.

\begin{table}[h]
\centering
\begin{tabular}{c | c |  c | c | c | c}
\hline \hline 
$z$ & GR (R1) & GR (R2[EdS]) & nDGP (R1) & nDGP(R2[EdS]) & nDGP (R2[EdS,UsA]) \\ 
\hline
0 & $0.31 \pm^{0.00}_{0.05} (0.17)$ & $0.29 \pm^{0.01}_{0.13}(0.16)$ & $0.37 \pm^{0.00}_{0.10} (0.17)$ & $0.35 \pm^{0.00}_{0.14}(0.16) $ & $0.38 \pm^{0.00}_{0.13}(0.16) $ \\ 
0.5 &  $0.18 \pm^{0.01}_{0.01} (0.36)$ & $0.18 \pm^{0.01}_{0.03}(0.23)$ & $0.20\pm^{0.01}_{0.03}(0.23)$ &$0.20\pm^{0.01}_{0.03}(0.23)$ & $0.21\pm^{0.01}_{0.03}(0.22)$ \\
1 &  $0.10 \pm^{0.01}_{0.01} (0.36)$ & $0.10\pm^{0.01}_{0.00} (0.36)$ & $0.11\pm^{0.01}_{0.00}(0.37)$ &$0.11\pm^{0.01}_{0.00}(0.36)$ & $0.12\pm^{0.01}_{0.00}(0.36)$\\
\hline
\end{tabular}
\caption{\footnotesize Best fit $\bar{c}_s^2 \equiv  c_s^2/ k_{\rm NL}^2$ in units of $(\unitsk)^{-2}$ with $k_{\rm fit}$ in units of $\unitsk$ in parentheses along with $1\sigma$ errors.}
\label{cs2list}
\end{table}

%
%
%
\subsection{Deviations from $\Lambda$CDM} \label{devsec}
\begin{figure}[h]
  \captionsetup[subfigure]{labelformat=empty}
  \centering
  \subfloat[]{\includegraphics[width=17 cm, height=8.1cm]{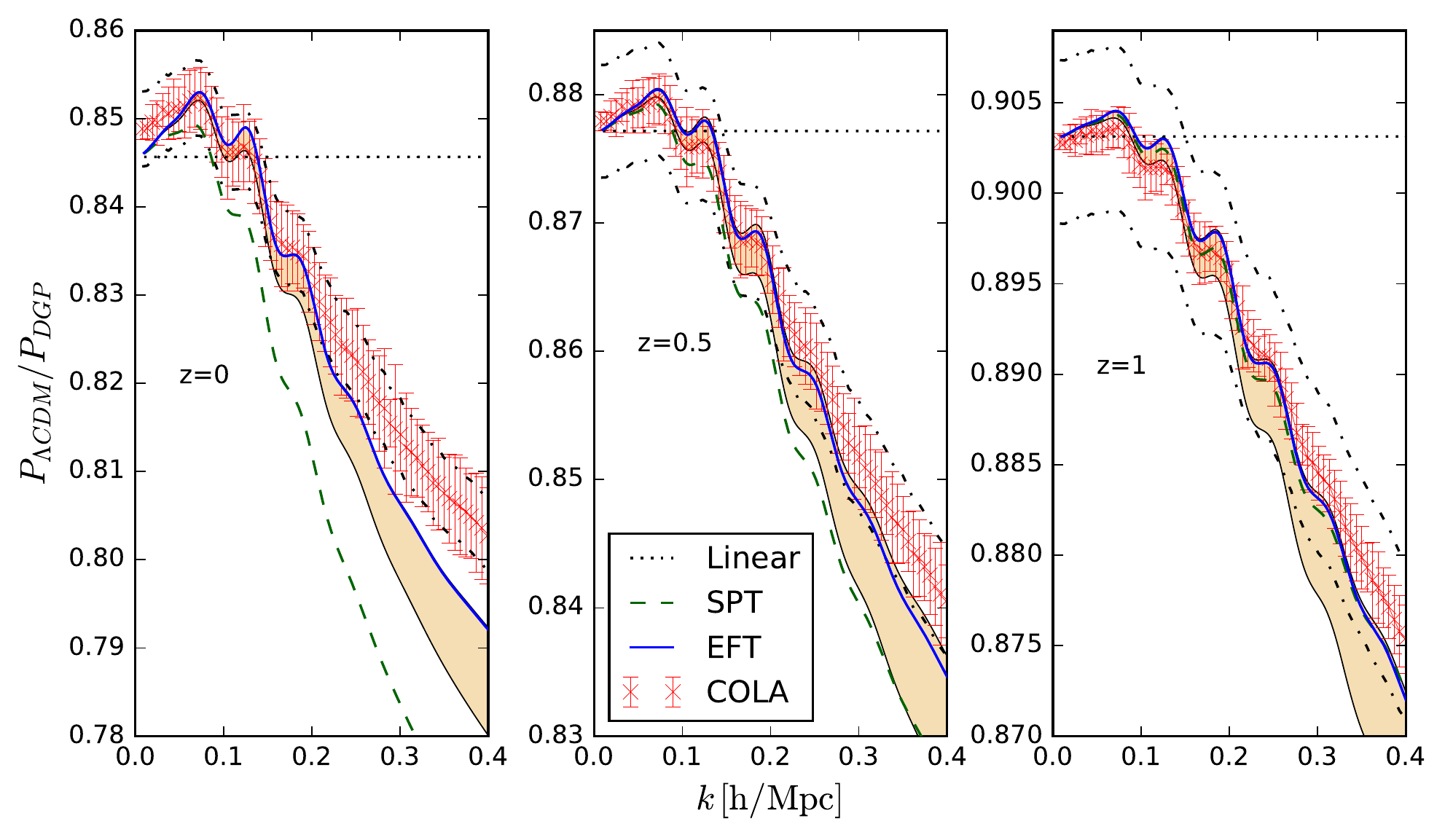}}
  \caption[CONVERGENCE ]{\footnotesize The ratio of the $\Lambda$CDM power spectrum to the nDGP power spectrum at $z=0$ (left), $z=0.5$ (center), and $z=1$ (right). Values of the fitted $\Delta \bar c_s^2$ are given in \tabref{cs2list2}.  The red crosses show the COLA measurements with $2\sigma$ error bars, $\sigma$ being the variance over 20 realizations, and the dot-dashed black curve is a constant $0.5 \%$ added systematic error explained in \secref{devsec}. The dotted black line gives the linear prediction, the dashed green curve gives the one-loop prediction in SPT, and the blue curve gives the one-loop EFTofLSS prediction using resummation scheme R1. The yellow band denotes the theoretical error, described in \appref{fittingproc}.}
\label{rateft}
\end{figure}

In this subsection, we consider the ratio of the $\Lambda$CDM and nDGP power spectra.  With the numerical data, the ratio is taken after each realization of the initial conditions (which are the same for $\Lambda$CDM and nDGP), so that the cosmic variance, which is related to averaging over many realizations of the initial conditions that all have the same power spectrum, is largely canceled.  The error on the ratio can then be determined by finding the variance over many simulations.  This variance is plotted in \figrefs{rateft}{rateft2}, and is much smaller than the variances plotted for the individual power spectra in previous plots.  Because of this, the ratio provides a precise prediction against which to test the computational methods at hand.  

To see what we will be sensitive to, consider expanding the EFTofLSS up to two loops, as
\begin{align}
P \simeq D^2 P_{11}^{\rm in} + P_{1\text{-loop}} - 2 ( 2 \pi) c_s^2 \frac{k^2}{\knl^2} D^2 P_{11}^{\rm in} + P_{2\text{-loop}}  \ , 
\end{align}
where  $P_{2\text{-loop}}$ denotes all the two-loop  terms, including counterterms.  Next, we take the ratio of this expression for $\Lambda$CDM and for nDGP to get, up to the same two-loop order,
\begin{align} \label{ratioformula}
\frac{P_{\Lambda\text{CDM}} }{P_{\rm nDGP}}&  \simeq \frac{D_{\Lambda\text{CDM}}^2}{D_{\rm nDGP}^2} \Bigg\{ 1 + \left( \frac{P_{1\text{-loop}}^{\Lambda\text{CDM}} }{D_{\Lambda\text{CDM}}^2 P_{11}^{\rm in} }-  \frac{P_{1\text{-loop}}^{  \rm nDGP} }{D_{\rm nDGP}^2 P_{11}^{\rm in} } \right)  - 2 ( 2 \pi) \frac{k^2}{\knl^2 }( c_{s,\Lambda\text{CDM}}^2 - c_{s,{\rm nDGP}}^2 )     \\
&\hspace{1in} +  \left( \frac{P_{2\text{-loop}}^{\Lambda\text{CDM}} }{D_{\Lambda\text{CDM}}^2 P_{11}^{\rm in} }-  \frac{P_{2\text{-loop}}^{  \rm nDGP} }{D_{\rm nDGP}^2 P_{11}^{\rm in} } \right)    +  2 ( 2 \pi)\frac{ k^2}{\knl^2}  \left( \frac{c_{s,\text{nDGP}}^2  P_{1\text{-loop}}^{\Lambda\text{CDM}} }{D_{\Lambda\text{CDM}}^2 P_{11}^{\rm in} }-  \frac{ c_{s,\Lambda\text{CDM}}^2  P_{1\text{-loop}}^{  \rm nDGP} }{D_{\rm nDGP}^2 P_{11}^{\rm in} } \right)      \Bigg\} \nonumber  \ .
\end{align}
In the one-loop EFT that we use, the first line of \eqn{ratioformula} is predicted and under control, while the second line is not.  For example, we do not compute $P_{2\text{-loop}}$, so this is absent in our computation.  The last term in the second line, proportional to $k^2 P_{1\text{-loop}}$, is included in our computation, but it is a higher order term, down by $k^2 / \knl^2 $ from the one-loop terms that we have included.

While this ratio is not an observable quantity, it does provide two very useful purposes because the error bars on the data are so small.  First, it is a powerful diagnostic to test how well the EFT and resummation schemes perform.  Second, as can be seen in \eqn{ratioformula}, the only combination of the counterterms that enters at the one-loop level is the difference $\Delta \bar c_s^2 \equiv \bar c_{s,{\rm nDGP}}^2- \bar c_{s, {\Lambda\text{CDM}}}^2$.  Thus, we can directly fit \eqn{ratioformula} to the data, treating $\Delta \bar c_s^2$ as the free parameter, and thus obtain a precise measurement of $\Delta \bar c_s^2$.  We will further discuss these two points below.

 The comparisons of various predictions and COLA measurements are shown in \figrefs{rateft}{rateft2}. The simulation measurements are shown as red crosses, now with the $2\sigma$ error bars where $\sigma$ is the variance of the ratio over the 20 runs.  The blue curves show the EFT  approach with exact time dependence and resummation scheme R1, with values of the fitted $\Delta \bar c_s^2$ given in \tabref{cs2list2}.  The green dashed line shows the one-loop SPT prediction and the dotted line is the linear prediction.

 The first thing to notice is that the $2 \sigma$ error bars on the data are extremely small: they are approximately $0.5\%$ at $z = 0$ and $0.2\%$ at $z = 1$.  While this is an incredible amount of precision, unfortunately we cannot take advantage of all of it in this study.  This is because the COLA simulations for nDGP have a systematic error associated with some approximations made within the code \citee{Winther:2017jof}.\footnote{In particular, the code uses a spherically symmetric approximation to determine the gravitational potential, and this affects the large scale clustering. This large scale effect is stronger at low $z$ because there is more non-linear structure formation. We refer the interested reader to \citee{Schmidt:2009sv}, where there is a discussion in Appendix A.}  These approximations allow the code to run quickly, and are generally less than $0.5\%$, which is more than enough accuracy to compute the individual power spectra since the cosmic variance does not go below about $1\%$.  However, we cannot be sure that this systematic error is not important in the ratio that we study here.  In order to take this into account, we add a $0.5\%$ systematic error, estimated from \citee{Winther:2017jof}, to the ratio in \figrefs{rateft}{rateft2} as a black dot-dashed line, and we use this error in our parameter fits.  This is still a very small error and allows us to make a precise measurement of $\Delta \bar c_s^2$; we present the even smaller variance in the data just to make the point that these measurements can be made more precise by running more accurate simulations.

 \begin{figure}[h]
  \captionsetup[subfigure]{labelformat=empty}
  \centering
  \subfloat[]{\includegraphics[width=17cm, height=8.1cm]{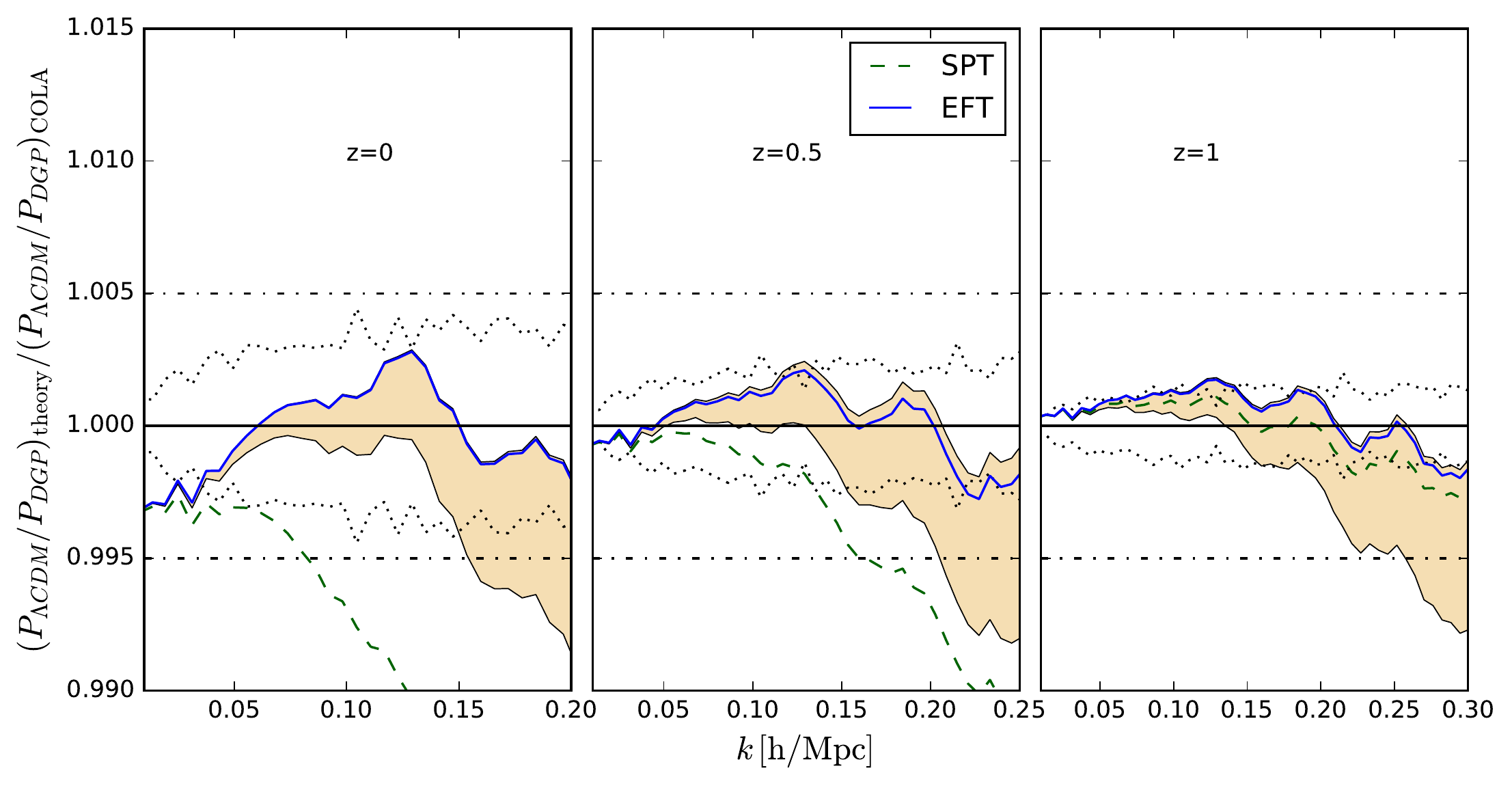}} 
  \caption[CONVERGENCE ]{ \footnotesize The ratio $(P_{\Lambda \text{CDM}} / P_{nDGP} )|_{\rm theory} / (P_{\Lambda \text{CDM}} / P_{nDGP} )|_{\rm COLA}$   at $z=0$ (left), $z=0.5$ (center), and $z=1$ (right). Values of the fitted $\Delta \bar c_s^2$ are given in \tabref{cs2list2}.   The dotted black curve shows the $2\sigma$ error bars on the COLA measurements, $\sigma$ being the variance over 20 realizations, and the dot-dashed black curve is a constant $0.5 \%$ added systematic error explained in \secref{devsec}.  The dashed green curve gives the one-loop prediction in SPT, and the blue curve gives the one-loop EFTofLSS prediction using resummation scheme R1. The yellow band denotes the theoretical error, described in \appref{fittingproc}.}
\label{rateft2}
\end{figure}

 In the end, we find a large improvement on the constraints for $\Delta \bar c_s^2$.  For example, if one were to use the individually measured parameters from \tabref{cs2list} for $\Omega_{\rm rc} = 0.438$ at $z = 0$, then, adding the errors in quadrature, one would find $(\unitsk)^2 \Delta \bar c_s^2 = (0.37 \pm^{0.00}_{0.10}) - (0.31 \pm^{0.00}_{0.05} ) \approx 0.06 \pm 0.06$, which is about a $100\%$ error.  This is to be contrasted with the measurement using the ratio, which is $(\unitsk)^2 \Delta \bar c_s^2= 0.067 \pm^{0.001}_{0.016} $, which is approximately a $13\%$ error.  Thus, we have shrunk the uncertainty by about a factor of 8.  Lastly, we  point out that while SPT and the EFT have a similar prediction at $z = 1$ for the ratio shown in \figref{rateft}, the predictions are much different for the individual power spectra shown in \figrefs{greft}{dgpeft}; this is simply due to the fact that $\Delta \bar c_s^2$ is very small at $z = 1$.  Next, we will use this fitting procedure to investigate the relationship between $\Delta  \bar c_s^2$ and the nDGP parameter $\Omega_{\rm rc}$.

%
%
%
\subsection{Effect of modified gravity on the speed of sound }
\label{soundspeed}

\begin{figure}[h]
  \captionsetup[subfigure]{labelformat=empty}
  \centering
  \subfloat[]{\includegraphics[width=17cm, height=8.1cm]{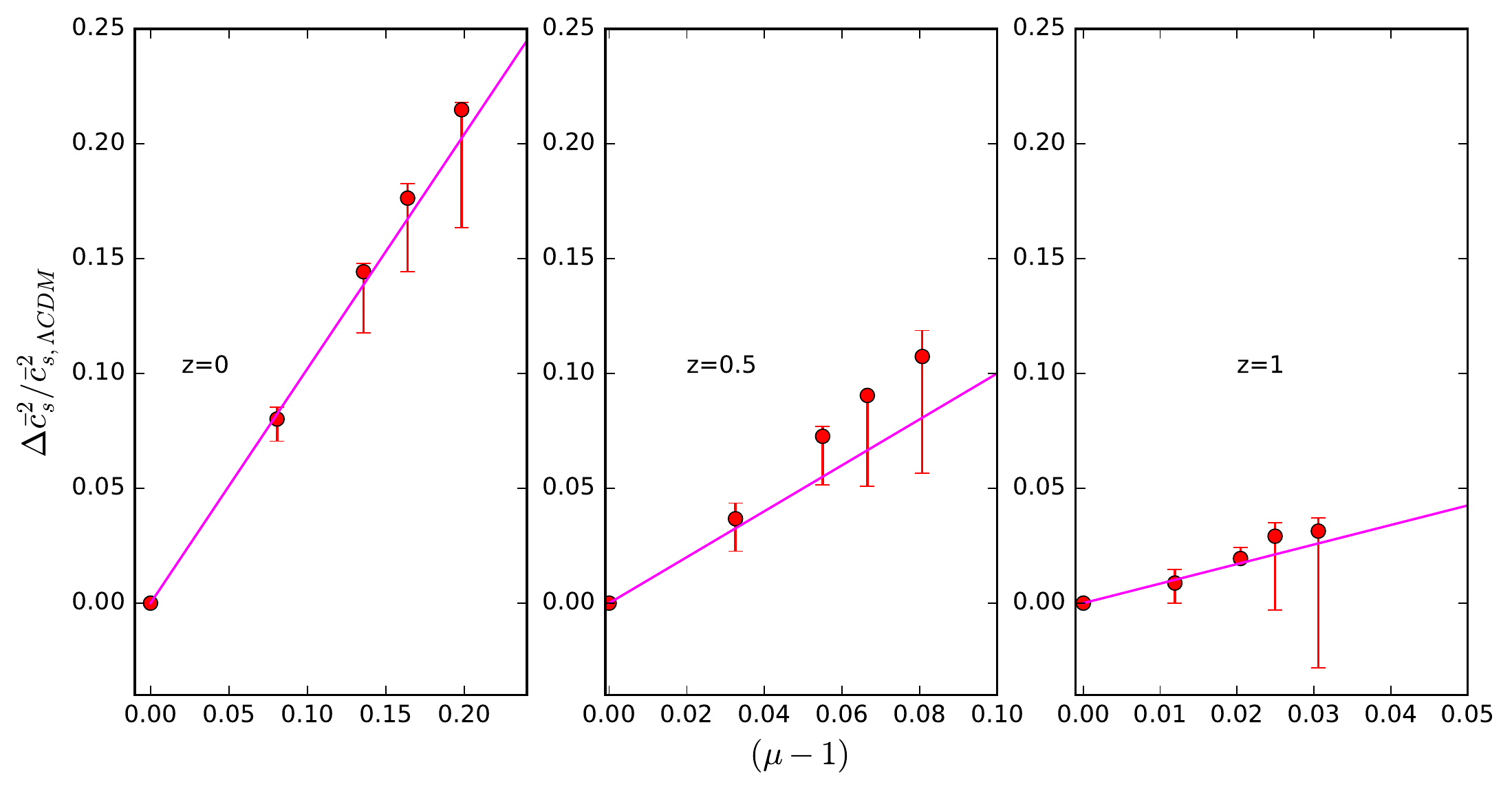}} 
  \caption[CONVERGENCE ]{ \footnotesize The relationship between $\Delta \bar {c}_s^2/\bar{c}_{s}^2(a,0) \equiv [\bar{c}_{s}^2(a,\Omega_{\rm rc})-\bar{c}_{s}^2(a,0)]/\bar{c}_{s}^2(a,0)$ and $\mu - 1$ (see \eqn{deltacs}) at $z=0$ (left), $z=0.5$ (centre) and $z=1$ (right). The errors shown are the $1\sigma$ errors coming from the $\chi^2$ fit, which is described in \appref{fittingproc}.  We have found the best fit value of the coefficients $A(z = 0) = 1.02$, $A(z = 0.5) = 1.00$, and $A(z = 1) = 0.85$.   }
\label{cs2vorc}
\end{figure}

In this subsection, we make use of the COLA simulations with various values of $\Omega_{\rm rc}$ between $0$ (the $\Lambda$CDM value) and $0.438$ to determine the dependence of $ \bar c_s^2 \equiv  c_s^2 / \knl^2$ on $\Omega_{\rm rc}$.  Specifically, for each of the values $\Omega_{\rm rc} \in \{ 0 , 0.035 , 0.135 , 0.235 ,  0.438 \}$, we have made 20 realizations with the same initial seeds and specifications as the simulations described in the previous subsection, for $z \in \{ 0 , 0.5 , 1\}$.  Then, for each value of $\omegarc$, we find the best fit and $1\sigma$ values for $\bar c_s^2$, as described in \appref{fittingproc}.  The results are shown in \figref{cs2vorc}.

In general, the  non-linear scale $\knl^2$ can change in modified gravity models due to two different effects \citee{Cusin:2017wjg}.  The first is due to modifications of the non-linear interactions (i.e. screening terms) in the fluid-like equations (i.e. $\mu_2 \sim 1 $ or $\mu_{22} \sim 1$ from \eqn{gamma2}).  The second is through the linear change $\mu$, which, even if $\mu_2\ll 1$ and $\mu_{22} \ll 1$, can change the non-linear scale through the standard $\Lambda$CDM non-linear vertices.  In the nDGP model that we consider here, the screening terms are extremely small (see \figref{usaandeds}), so the latter effect is the important one.  

As shown in \citee{Cusin:2017wjg} for this case for general modified gravity models, the non-linear scale $\knl^2$ is expected to change by something proportional to $\mu ( a ) -1$ (when $\mu(a) -1$ can be treated perturbatively in the linear equations of motion), i.e. proportional to the change in the effective Newton's constant, $\Delta G_N / G_N$.  Because $ c_s^2$ and $\knl^2$ are degenerate at one loop, we can view this as a change in $\bar c_s^2$, so we expect
\be \label{deltacs}
\frac{\bar c_s^2 ( a;\omegarc ) - \bar c_s^2 (a; 0 ) }{\bar c_s^2 ( a;0)} =  A(a) \left( \mu( a ) -1 \right)  + \mathcal{O}\left(  \mu( a ) -1  \right)^2 = \frac{ A ( a ) \sqrt{\omegarc}}{3 \left( \sqrt{\omegarc} + \frac{H}{H_0} \left( 1 + \frac{ a H' }{3 H} \right) \right)}  + \mathcal{O}(\omegarc )  \ ,
\ee
where $A(a)$ is a time dependent function of order unity.\footnote{The time dependence is estimated in \citee{Cusin:2017wjg} for a scaling universe but, in this work, we simply fit the coefficient at each redshift.  In general, the time dependence of the counterterms is not predictable within the EFTofLSS.  Although there are well motivated estimates for the time dependence that one can use to define an advantageous parameterization, the time dependence should always be considered free. Thus, at this stage, we do not find it too enlightening to estimate the time dependence of $A(a)$, since we have already started with a well motivated time dependence in \eqn{deltacs}.  }  

To test the estimate \eqref{deltacs}, we also plot in \figref{cs2vorc} the functional form in \eqn{deltacs} and find the best fit value of $A(a)$.  We found the best fit values  $A(z = 0) = 1.02$, $A(z = 0.5) = 1.00$, and $A(z = 1) = 0.85$, consistent with our expectation that $A \sim \mathcal{O}(1)$.  From \figref{cs2vorc} we see that the analytic expression \eqn{deltacs} does very well in capturing the dependency of $\bar c_s^2$ on $\Omega_{\rm rc}$. 

\begin{table}[h]
\centering
\begin{tabular}{|c | c |  c | c |}
\hline \hline 
$\Omega_{\rm rc}$ &$z=0$ & $z=0.5$ & $z=1$ \\ 
\hline
0.035 & $ 0.025 \pm^{0.002}_{0.003} (0.38) $ & $ 0.007 \pm^{0.001}_{0.003}$ (0.40) & $ 0.001 \pm^{0.000}_{0.001}$ (0.60)   \\ 
0.135 & $ 0.045 \pm^{0.000}_{0.008} (0.26) $ & $ 0.013 \pm^{0.001}_{0.004}$ (0.35) & $ 0.002 \pm^{0.001}_{0.002}$ (0.48)  \\ 
0.235 & $ 0.055 \pm^{0.002}_{0.010} (0.23) $ & $ 0.016 \pm^{0.000}_{0.007}$ (0.28) & $  0.003 \pm^{0.001}_{0.003}$ (0.37)  \\ 
0.438 & $ 0.067 \pm^{0.001}_{0.016} (0.20) $ & $ 0.019 \pm^{0.002}_{0.010}$ (0.23) & $  0.003 \pm^{0.001}_{0.006}$ (0.23) \\
\hline
$\Lambda$CDM & 0.31 & 0.18 & 0.10\\
\hline
\end{tabular}
\caption{\footnotesize Best fit $\Delta \bar{c}_s^2 \equiv \bar{c}_{s}^2(z,\Omega_{\rm rc})-\bar{c}_{s}^2(z,0)$ in units of $(\unitsk)^{-2}$ with $k_{\rm fit}$ in units of $\unitsk$ in parentheses along with $1\sigma$ errors.  The last row is the corresponding value of $\bar c_s^2 ( z , 0)$ from the individual fits in \tabref{cs2list}, for reference.}
\label{cs2list2}
\end{table}

%
%
%
\section{Summary and conclusions} \label{concsec}

We have compared the one-loop prediction for the dark matter power spectrum computed in the framework of the effective field theory of Large-Scale Structure, with non-linear data sets of 20 COLA simulations, in the case where gravity is described, in the quasi-static non-relativistic approximation, either by general relativity or by the normal branch of the DGP model. 

The most theoretically justified analytical derivation of the one-loop power spectrum includes the exact calculation of the Green's functions to solve for the time dependence  of perturbations, briefly reviewed in Sec.~\ref{theorysec}, and the IR resummation of bulk modes, given in  \eqn{r1} (Sec.~\ref{resummationsec}).
We compare the results of this calculation to three  approximations commonly used for the calculation of the one-loop power spectrum in standard perturbation theory: 
the Einstein de Sitter (EdS) approximation,
where one assumes that the $\Lambda$CDM part of the nDGP power spectrum has simple EdS time dependence (see \eqn{edsapprox}), the unscreened approximation (UsA) where one ignores the screening terms in the nDGP power spectrum, and the IR resummation scheme in \eqn{r2}, which is based on splitting the power spectrum into a wiggle and no-wiggle part.  

First, in Sec.~\ref{sub1}, we  quantified the accuracy of the UsA and the EdS approximations in standard perturbation theory (SPT). We found that, even for large values of $\Omega_{\rm rc}$ (the constant paramaterizing the deviations from $\Lambda$CDM in nDGP, see Sec.~\eqref{sub_setup}), the effect of these approximations is sub percent below $k = 0.2 \unitsk$ (see \figref{usaandeds}). The UsA approximation gives a systematic increase in power that grows with $k$, which is maximized for $\Omega_{\rm rc} \approx 1$ and reaches approximately $1\%$ at $k \approx 0.2 \unitsk$ at $z=0$. On the other hand, the EdS approximation  gives a systematic decrease in power but is a much smaller effect. This is sub $0.25\%$ for $z=0.5$ and $z=1$; the largest deviation is for nDGP at $z = 0$, which reaches approximately $1\%$ at $k \approx 0 .2 \unitsk$.  

Remaining in SPT, in Sec.~\ref{comparisonsec1} (see \figrefs{grspt}{dgpspt}) we  compared the two IR resummation schemes and found that they are consistent to within $1\%$ for $k \leq 0.2 \unitsk$.  These results can be relevant for future data comparisons. Indeed, it was found that such sub-percent deviations may still be impactful in the context of stage IV spectroscopic surveys \citee{Bose:2017myh}. 

Next, in Sec.~\ref{comparisonsec2}, we looked at the one-loop EFT power spectra.  An important feature of the EFT computation was the inclusion of theoretical errors, which helped define the overall reach of the theory.  Given the errors on the data, for $\Lambda$CDM we found $k_{\rm reach} ( z = 0 ) \approx 0.14\unitsk$, $\kreach ( z = 0.5 ) \approx 0.30\unitsk$, and $\kreach(z=1) \approx 0.34 \unitsk$, and for nDGP we found, $k_{\rm reach} ( z = 0 ) \approx 0.14\unitsk$, $\kreach ( z = 0.5 ) \approx 0.35 \unitsk$, and $\kreach(z=1) \approx 0.36 \unitsk$.  The fitted values of the counterterms can be found in \tabref{cs2list}.  We also compared the various approximations amongst themselves, and found that the EdS and UsA approximations amounted to less than $1\%$ deviations over the scales of interest. This suggests that using such approximations within the EFTofLSS, especially at higher $z$, can still give valid results. Of course this depends on the model of gravity. In certain scenarios, screening may be more important on non-linear scales \cite{Fasiello:2017bot} and in $f(R)$ theories (see for example \citee{Sotiriou:2008rp,Burrage:2017qrf} for reviews) separating time- and $k$-dependence can become completely invalid.

Moreover, in Sec.~\ref{devsec} we used the ratio of $\Lambda$CDM and nDGP power spectra to give a precision test of our theories.  In particular, in the EFTofLSS, we directly fit the difference in the speeds of sound $\Delta \bar{c}_s^2$ to the ratio data.  For example, for $\Omega_{\rm rc} = 0.438$ at $z = 0$, this gave us error bars on $\Delta \bar{c}_s^2$ that were about a factor of 8 smaller than what we would have obtained by simply taking the difference of the individually fit parameters.  

Next, in Sec.~\ref{soundspeed}, we showed by fitting to COLA simulations with different $\Omega_{\rm rc}$, that $\Delta \bar{c}_s^2$ is essentially proportional to $\mu( a) -1$, i.e. the change in the effective Newton's constant, allowing for an analytic prediction of the sound speed parameter dependence on $\Omega_{\rm rc}$.  When combined with an advantageous parameterization for the redshift dependence of the proportionality factor, such a prediction would be very powerful in theory-data comparisons when using the EFTofLSS framework. For example, Ref.~\cite{Foreman:2015uva}  investigates the redshift dependence of $\bar{c}_s^2$ and finds that it is a function of the effective tilt of the linear power spectrum and its derivatives at a scale $k_{\rm ren}$, where the two-loop computation becomes sizeable.  Combining these ingredients could offer a potential means of putting strong priors on the free parameters of the EFTofLSS, drastically enhancing its constraining power. We leave a full investigation into these dependencies of $\bar c_s^2$ in theories beyond GR for another work.

Our results are of particular relevance for high precision lensing measurements such as LSST \citee{Chang:2013xja}, where small theoretical inaccuracies can lead to biased constraints of gravity. The EFTofLSS offers a significant improvement over SPT in terms of range of scales, but it is still unknown, because of the addition of free parameters, by how much this will improve parameter constraints.  For this reason, analytic forms for the gravity dependence of sound speed parameters will be very important in strengthening constraints as well as optimizing parameter inference pipelines. In a future work, we will extend this study to redshift space  in preparation for very large volume spectroscopic surveys that are set to come online in the near future.

\section*{Acknowledgments}
\noindent M.L. would like to thank S. Foreman for useful discussions regarding the fitting procedure used here.  M.L. and F.V. thank F.~Schmidt and A.~Barreira for interesting discussions at the start of this project.  B.B. is supported by the University of Portsmouth. K.K. is supported by the UK Science and Technologies Facilities Council grants ST/N000668/1.  K.K. and H.A.W. have received funding from the European Research Council (ERC) under the European Union's Horizon 2020 research and innovation programme (grant agreement No.646702 ``CosTesGrav'').  M.L.~acknowledges financial support from the Enhanced Eurotalents fellowship, a Marie Sklodowska-Curie Actions Programme. F.V.~acknowledges financial support from ``Programme National de Cosmologie and Galaxies'' (PNCG) of CNRS/INSU, France and  the French Agence Nationale de la Recherche under Grant ANR-12-BS05-0002. 
%
%
%
%
%

\appendix

\newpage
\appendix

%
%
%
%
\section{Green's functions} \label{exacttimeapp}

In this appendix, we provide the explicit formulae for the Green's functions used in Sec.~\ref{linearsolsec} and throughout this work.  We use the same notation as in \citee{Cusin:2017wjg}, so that the formulae there in App. C II - C IV are directly applicable to this paper.  Using the perturbative expansion \eqref{dtGreen} and \eqref{dtGreen2} in the continuity and Euler equations \eqref{eq:Perturb1} and \eqref{eq:Perturb2}, we find that the four Green's functions are specified by the following equations
 \begin{align}
 &a \frac{d G^{\delta}_{\sigma}(a,\ta)}{da}   -  G^{\Theta}_{\sigma}(a,\ta)=\lambda_{\sigma}\delta(a-\ta) \ ,  \label{Green} \\
 &a \frac{d G^{\Theta}_{\sigma}(a,\ta)}{da}    + \left( 1 + \frac{ a \cH'(a)}{\cH(a)} \right)  G^{\Theta}_{\sigma}(a,\ta) -  \mu ( a ) \frac{3 \, \omegam (a )}{2   }  G^{\delta}_{\sigma}(a,\ta)    =(1-\lambda_{\sigma})\delta(a-\ta) \ ,
 \end{align}
where $\sigma=\{1,2\}$, $\lambda_1=1$, $ \lambda_2=0$, and $\delta(a-\ta)$ is the Dirac delta function.  The retarded Green's functions satisfy the boundary conditions 
\be
\begin{split}  
& G^{\delta}_\sigma(a,\tilde a)  =  0 \quad \quad  \text{and}   \quad\quad G^{\Theta}_\sigma(a, \tilde a)=0  \quad \quad \text{for} \quad \quad \tilde a > a  \ , \\
 &G^\delta_\sigma ( \tilde a , \tilde a ) = \frac{\lambda_\sigma}{\tilde a}  \quad \hspace{.06in} \text{and} \hspace{.2in}  \quad G^{\Theta}_{\sigma} ( \tilde a  , \tilde a ) = \frac{(1 - \lambda_\sigma)}{\tilde a}  . \label{bound2}
\end{split}
\ee

We can then construct the Green's functions in the usual way using the linear solutions and the Heaviside step function, $\Theta_{\rm H} (a-\tilde a)$, and imposing the boundary conditions \eqref{bound2}.  This gives 
\begin{align}
&G^{\delta}_1(a,\ta) = \frac{1}{\ta W(\ta)}\bigg(\frac{d D_{-}(\ta)}{d\ta}  D_{+}(a)-\frac{d D_{+}(\ta)}{d\ta}D_{-}(a)\bigg)\Theta_{\rm H}(a-\ta)  \label{gdelta} \ , \\
&G^{\delta}_2(a,\ta)=  -  \frac{1}{ \ta^2 W(\ta)}    \bigg(  D_{-}(\ta)D_{+}(a)  - D_{+}(\ta)D_{-}(a)    \bigg)\Theta_{\rm H}(a-\ta)  \label{gdelta2} \ , \\
&G^{\Theta}_1(a,\ta)=   \frac{1}{ \ta W(\ta)}   \bigg(\frac{d D_{-}(\ta)}{d\ta}   \frac{a \, d D_{+}(a)}{d a}-\frac{d D_{+}(\ta)}{d\ta}     \frac{   a \, d D_{-}(a)}{d a}\bigg)\Theta_{\rm H}(a-\ta) \ ,\\
&G^{\Theta}_2(a,\ta)  =  - \frac{1}{ \ta^2 W(\ta)}    \bigg(  D_{-}(\ta)\frac{a \, d D_{+}(a)}{d a}  -  D_{+}(\ta)\frac{a\, d D_{-}(a)}{d a}     \bigg)\Theta_{\rm H}(a-\ta) \ ,   \label{gtheta}
\end{align}
where $W(\ta)$ is the Wronskian of $D_+$ and $D_-$, i.e. $W(\ta)= D_{-}' (\ta) D_{+}(\ta)-  D_{+}'(\ta) D_{-}(\ta) $.

%
%
%
%
\section{IR-resummation details} \label{irresumapp}

In this appendix, we present some details for doing the IR-resummation presented in \sect{resummationsec}.  First, we introduce the following notation.  We use a double bar $g(k;a)||_n$ to mean that the quantity $g(k;a)$ is expanded up to $n$-th order in all of the parameters $\epsilon_{s <}$, $\epsilon_{ \delta < }$, and $\epsilon_{s >}$: this is simply the Eulerian expansion.  We use a single bar $g(k;a)|_n$ to mean that $g(k;a)$ is expanded to up $n$-th order in $\epsilon_{ \delta < }$ and $\epsilon_{s >}$, but that $\epsilon_{s<}$ has been resummed: this is the result of the IR-resummation.  Finally, for the power spectrum, we use a subscript like in $P(k ; a )_j$ to mean that we take the $j$-th loop-order piece of the power spectrum in Eulerian perturbation theory.

The formula for the resummed power spectrum at order $N$ is\footnote{In these formulae, the double bar $||$ that appears in the subscripts of $M$ and $F$ is just a part of the name of these functions and does not mean that $M$ and $F$ are themselves expanded.  The quantity that is actually expanded is $K_0^{-1}$ in \eqn{expandf}.}
\begin{align} \label{r15}
P ( k; a ) \Big|_N & = \sum_{j = 0}^N  \int \frac{d k' \, k'^2}{2 \pi^2} M_{||_{N-j}} ( k , k' ; a ) P ( k' ;a )_j \ ,  \\
M_{||_{N-j}} ( k , k' ; a ) & = \int d q \,\,j_{0 } ( k' q ) \, q^2    \int d^2 \hat q \, e^{- i \qvec \cdot \kvec }     F_{||_{N-j}} ( \kvec , \qvec ;a  )  \ ,  \label{sbt}
\end{align}
where $j_0(x) \equiv (\sin x ) / x$ is the zeroth spherical Bessel function, and we will discuss the other ingredients, as well as the computational strategy, below.

The function $ F_{||_{N-j}} ( \kvec , \qvec ;a  )$ contains all of the information about the long-wavelength displacements, and is given by 
\be \label{expandf}
F_{||_{N-j}}   \left(\kvec ,\qvec  ;   a       \right)   =    K_{0} \left( \kvec , \qvec  ; a   \right) \cdot \left( \left.\left.K_{0}^{-1}\left( \kvec ,\qvec ; a  \right)\right|\right|_{N-j}  \right) \ ,
\ee
where \citee{Senatore:2014via}
\begin{align} \label{k01}
K_0 ( \kvec , \qvec ; a) &  = \exp \left\{ - \frac{k^2}{2}  \left( X_1 ( q ; a  )  + Y_1(q ; a ) ( \kdotq )^2 \right)  \right\} \ , 
\end{align}
with 
\begin{align} \nonumber
 X_1(q;a) & =\frac{1}{2\pi^2}\int_0^{+\infty} dk\;  \exp\left[- \frac{k^2}{\Lambda_{\rm IR}^2}\right]\;P_{11}(k;a) \left[\frac{2}{3}-2\,\frac{j_1(k q)}{k q}\right]\  \\
 Y_1(q;a) & =\frac{1}{2\pi^2}\int_0^{+\infty} dk\;\exp\left[- \frac{k^2}{\Lambda_{\rm IR}^2}\right]\; P_{11}(k;a) \left[- 2\, j_0(k q)+6\,\frac{j_1(k q)}{k q}\right]\ 
\end{align}
and $j_0$ and $j_1$ are the zeroth and first spherical Bessel functions respectively.  The functions $X_1$ and $Y_1$ come from the power spectrum of the linear displacement field, $\bfs_1 ( \kvec ; a )$, which is given by $- i \kvec \cdot \bfs_1 = \delta_1$ for $\kvec \neq 0$ .  Here, $\Lambda_{\rm IR}$ is the IR scale up to which we resum the linear IR modes.  This cutoff and the choice in \eqn{k01} to keep the linear modes non-perturbative both serve to define a new expansion parameter $\tilde \epsilon_{s<} $, such that $\tilde \epsilon_{s<} \ll 1 \lesssim \epsilon_{s<}$, in terms of which the perturbative expansion of the IR displacements is now done.  Taking $\Lambda_{\rm IR}$ too high would mean that we include some uncontrolled UV modes, but this mistake would be at the next order in $\epsilon_{\delta <}$ and would be recovered order by order in the loop expansion.  In practice, we use $\Lambda_{\rm IR} = 0.066 \unitsk$.  

Although this method has the advantage that one controllably treats the IR displacements, the practical computation can be challenging.  Although not too demanding at one-loop in real space, it can become more complex at higher loops and in redshift space.  Because of this, \citee{Lewandowski:2015ziq} showed that one can expand \eqn{k01} (and the analogous expression in redshift space) in $k^2 Y_1 ( q ; a )$ to some desired order, in which case the angular integral in \eqn{sbt} becomes analytic.  Indeed, this is how we have actually evaluated the resummation in this work, although we refer the reader to \citee{Lewandowski:2015ziq} for details.

\section{Fitting procedure and theory errors} \label{fittingproc}

Here we give details on our fitting procedure for EFTofLSS parameter $ \bar c_s^2$. In this scheme, one minimizes the $\chi^2$ for a given $k_{\rm max}$
\begin{equation}
\chi^2(k_{\rm max}, \bar c_s^2) = \sum_{i={\rm min}}^{{\rm max}} \frac{ \left[P_{\rm COLA}(k_i)-P_{\rm EFT}(k_i, \bar c_s^2)\right]^2}{\sigma_i^2},
\end{equation}
where $\sigma_i$ is the variance of the data at point $i$ over the 20 realizations, to determine a best fit value $\bar c_s^2 ( k_{\rm max})$. Then, one increases the value of $k_{\rm max}$ and does the same $\chi^2$ minimization again for each value of $k_{\rm max}$.  Using this, one determines a $k_{\rm max}$ such that the value of $\bar c_s^2$ remains within some specified confidence interval\footnote{For example, \cite{Foreman:2015lca} used a $2\sigma$ criterion.} of the best fit values at lower $k_{\rm max}$.  The value of $k$ where the central value drifts outside of the error bars of lower wavenumbers is called $k_{\rm fit}$.  This procedure was introduced in \cite{Foreman:2015lca} for fitting the two-loop dark-matter power spectrum to help avoid over-fitting the data.  The main idea is that when the fitted value of $\bar c_s^2$ goes outside of the error bars of the values obtained with a lower $k_{\rm max}$, the $\chi^2$ minimization is trying to compensate for the fact that we have left off higher order terms.  This signals that we are outside the regime of validity of the one-loop computation.

 \begin{figure}[h]
  \captionsetup[subfigure]{labelformat=empty}
  \centering
  \subfloat[]{\includegraphics[width=17cm, height=8.1cm]{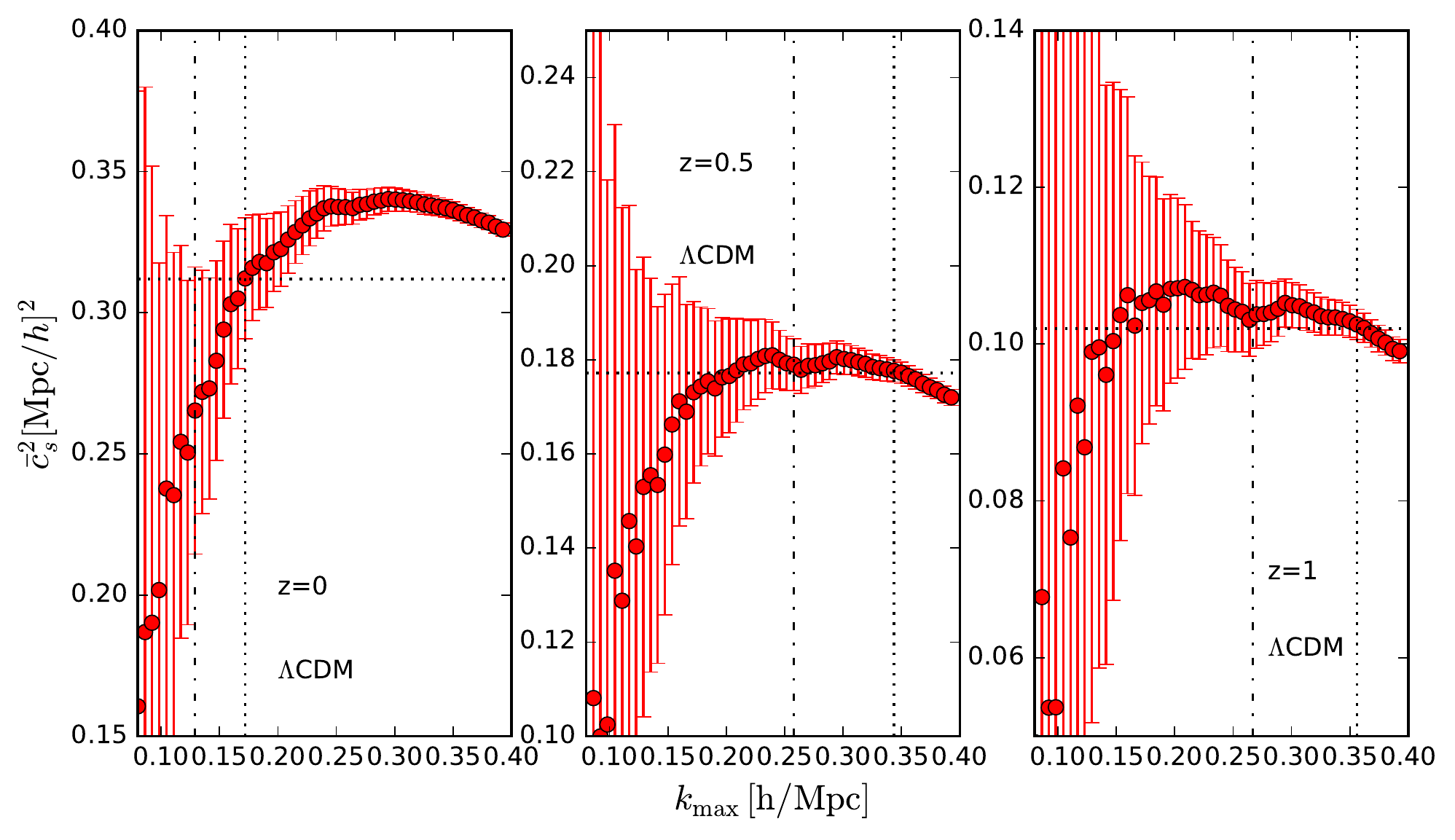}} 
  \caption[CONVERGENCE ]{\footnotesize The best fit value of $\bar{c}_s^2= c_s^2/ k_{\rm NL}^2$  as a function of $k_{\rm max}$ for $\Lambda$CDM at $z=0$ (left), $z=0.5$ (centre) and $z=1$ (right) with the associated $1\sigma$ confidence intervals. The fits are done using the R1 resummation method. The vertical dotted line is $k_{\rm fit}$, the vertical dot-dashed line is $0.75 k_{\rm fit}$, and the horizontal dotted line is the best fit value of $\bar c_s^2$.  }
\label{kfits1}
\end{figure}

 \begin{figure}[h]
  \captionsetup[subfigure]{labelformat=empty}
  \centering
  \subfloat[]{\includegraphics[width=17cm, height=8.1cm]{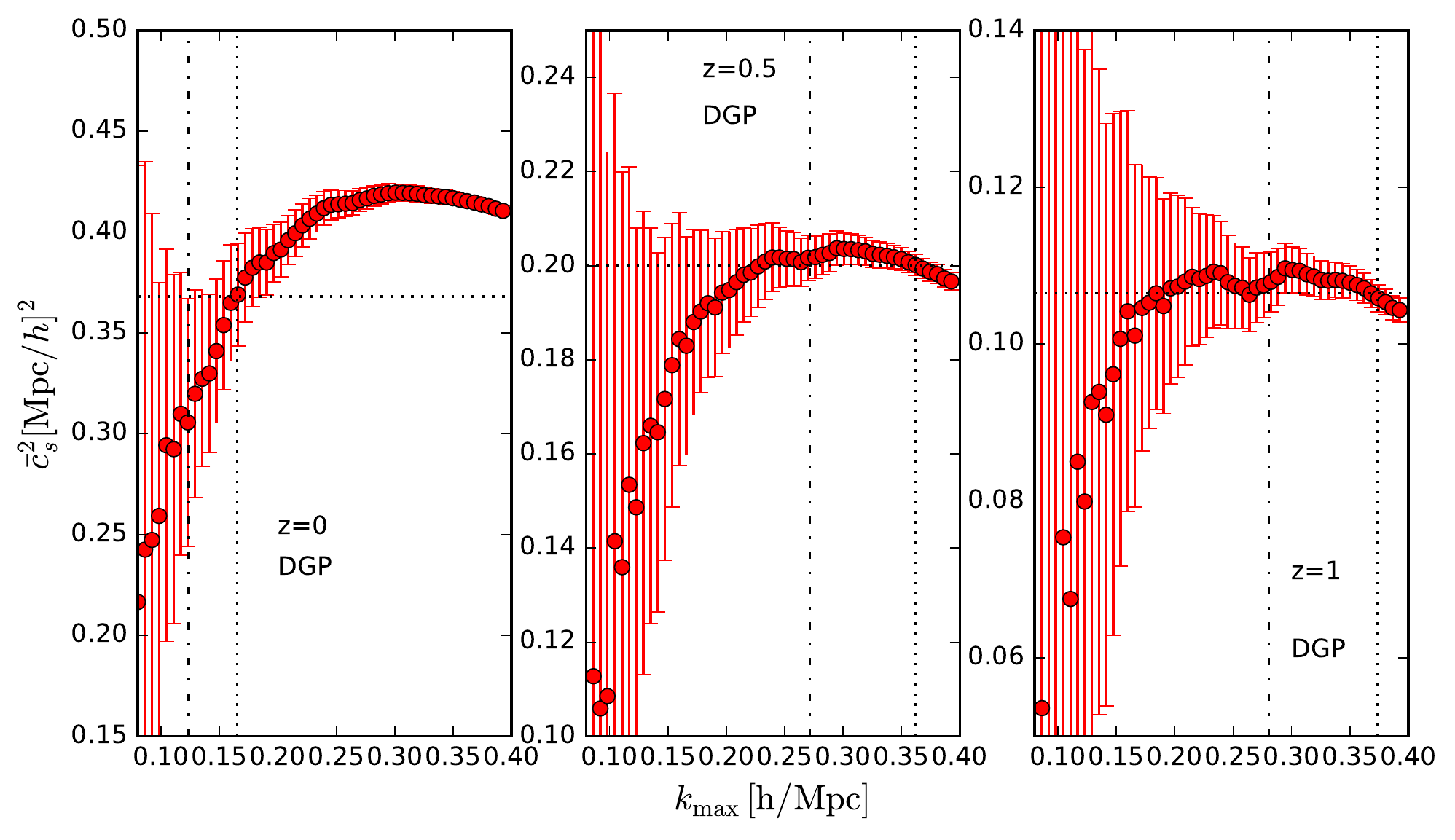}} 
  \caption[CONVERGENCE ]{\footnotesize The best fit value of $\bar{c}_s^2= c_s^2/ k_{\rm NL}^2$  as a function of $k_{\rm max}$ for nDGP ($\Omega_{\rm rc} = 0.438$) at $z=0$ (left), $z=0.5$ (centre) and $z=1$ (right) with the associated $1\sigma$ confidence intervals. The fits are done using the R1 resummation method. The vertical dotted line is $k_{\rm fit}$, the vertical dot-dashed line is $0.75 k_{\rm fit}$, and the horizontal dotted line is the best fit value of $\bar c_s^2$. }
\label{kfits2}
\end{figure}
\figref{kfits1} and \figref{kfits2} show the best fit $\bar{c}_s^2 \equiv  c_s^2/ k_{\rm NL}^2$ against $k_{\rm max}$ for $z=0$, $0.5$, and $1$ for $\Lambda$CDM and nDGP respectively, along with the $1\sigma$ errors.  For example, let us look at the $\Lambda$CDM fits in \figref{kfits1}.  At $z = 0$, the $k_{\rm max}$ where the central value leaves the lower error bars is quite easy to see: this occurs near $k_{\rm fit} \approx 0.17 \unitsk$, where the central value exits the error bars of the ``dip'' near $k \approx 0.13 \unitsk$.   For $ z = 0.5$ and $ z = 1$, because the theory fits much better at higher redshift, the situation looks slightly different, but the idea is the same.  In both plots, the value of $\bar c_s^2$ is converging, and every central value is within the error bars of the lower $k_{\rm max}$ fits.  However, in each plot, at some point the fits stop converging, and the central value starts to drift slightly.  This occurs near $k_{\rm fit} \approx 0 .36 \unitsk$ for both $ z = 0.5$ and $z = 1$.  Fit parameters are summarized in \tabref{cs2list}.

 \begin{figure}[h]
  \captionsetup[subfigure]{labelformat=empty}
  \centering
  \subfloat[]{\includegraphics[width=17cm, height=8.1cm]{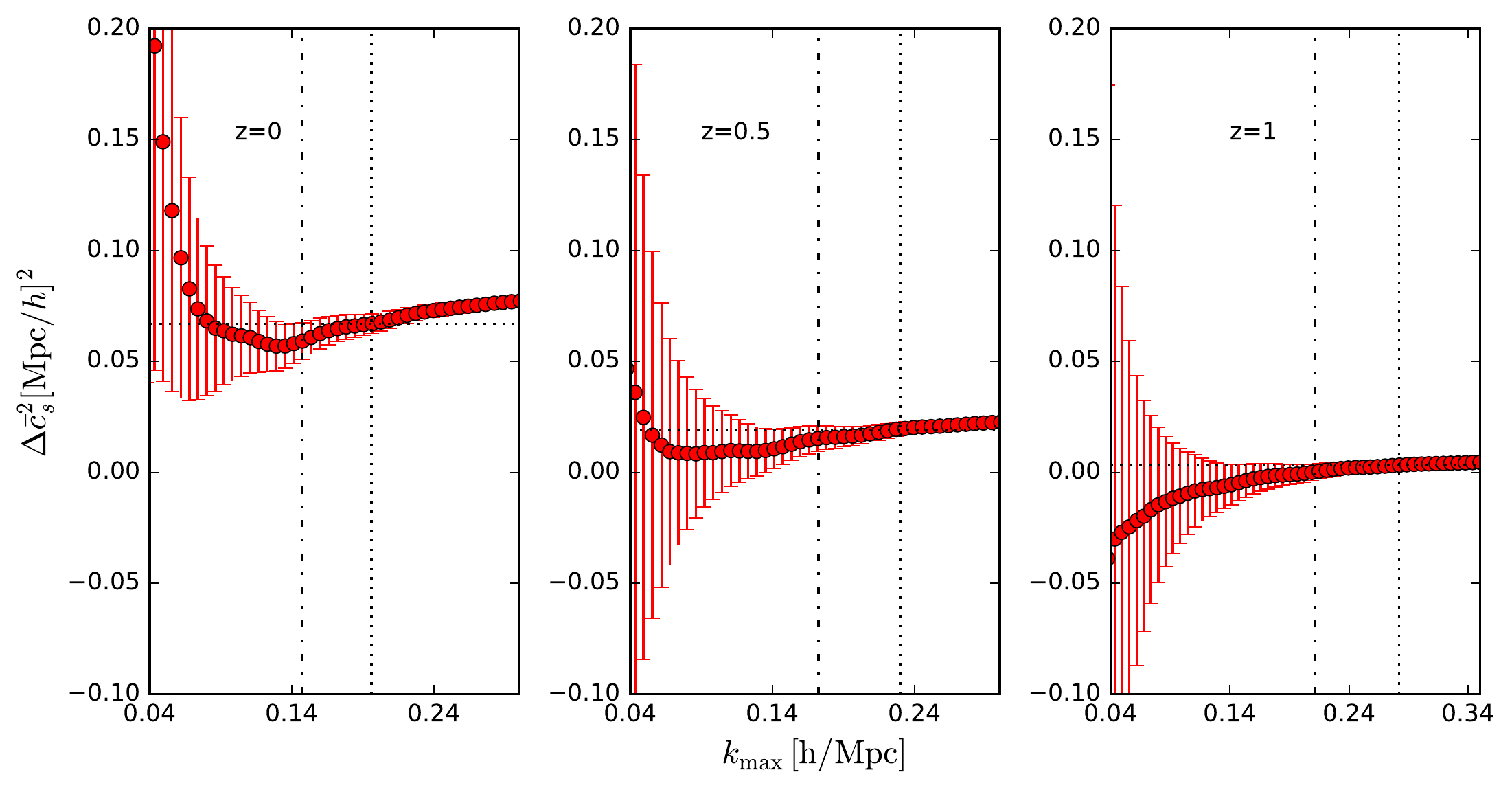}} 
  \caption[CONVERGENCE ]{\footnotesize The best fit value of $\Delta \bar{c}_s^2=\bar{c}_{s,{\rm nDGP}}^2-\bar{c}_{s,{\Lambda\text{CDM}}}^2$  as a function of $k_{\rm max}$ at $z=0$ (left), $z=0.5$ (centre) and $z=1$ (right) with the associated $1\sigma$ confidence intervals. The fits are done using the R1 resummation method. The vertical dotted line is $k_{\rm fit}$, the vertical dot-dashed line is $0.75 k_{\rm fit}$, and the horizontal dotted line is the best fit value of $\Delta \bar c_s^2$.  }
\label{kfits3}
\end{figure}
Similarly, \figref{kfits3} shows the best fit $\Delta \bar{c}_s^2 \equiv [\bar{c}_{s}^2(z,\Omega_{\rm rc})-\bar{c}_{s}^2(z,0)]/\bar{c}_{s}^2(z,0)$ when comparing to the ratio of the COLA data. In this fitting we fix $\bar{c}_s^2(z,0)$ to the best fit value obtained by fitting just the $\Lambda$CDM spectrum (see Tab.~\ref{cs2list}) and then fit the difference. We have checked that changing $\bar{c}_s^2(z,0)$ does not change the fit of $\Delta \bar{c}_s^2$.  Indeed, one can see from \eqn{ratioformula} that at one loop, the only relevant parameter is $\Delta \bar c_s^2$.  The fit parameters are summarized in \tabref{cs2list2}.

Finally, we discuss the theoretical error associated with the EFT computation.  There are two general sources of theoretical error: the finite truncation of the EFT expansion, and the uncertainty in the fitting procedure described above.  The former is of a theoretical nature and serves as a lower bound on the uncertainty in the computation; one can never claim to have a computation that is more precise than the first loop level that has been omitted from the computation.  On the other hand, the uncertainty due to the fitting procedure is driven by the data itself, including the size of its error bars.  Thus, one should estimate the uncertainties due to both of these effects, and use the largest as the theoretical error.  Next, we will do just that.

\begin{figure}[h]
  \captionsetup[subfigure]{labelformat=empty}
  \centering
  \subfloat[]{\includegraphics[width=14cm]{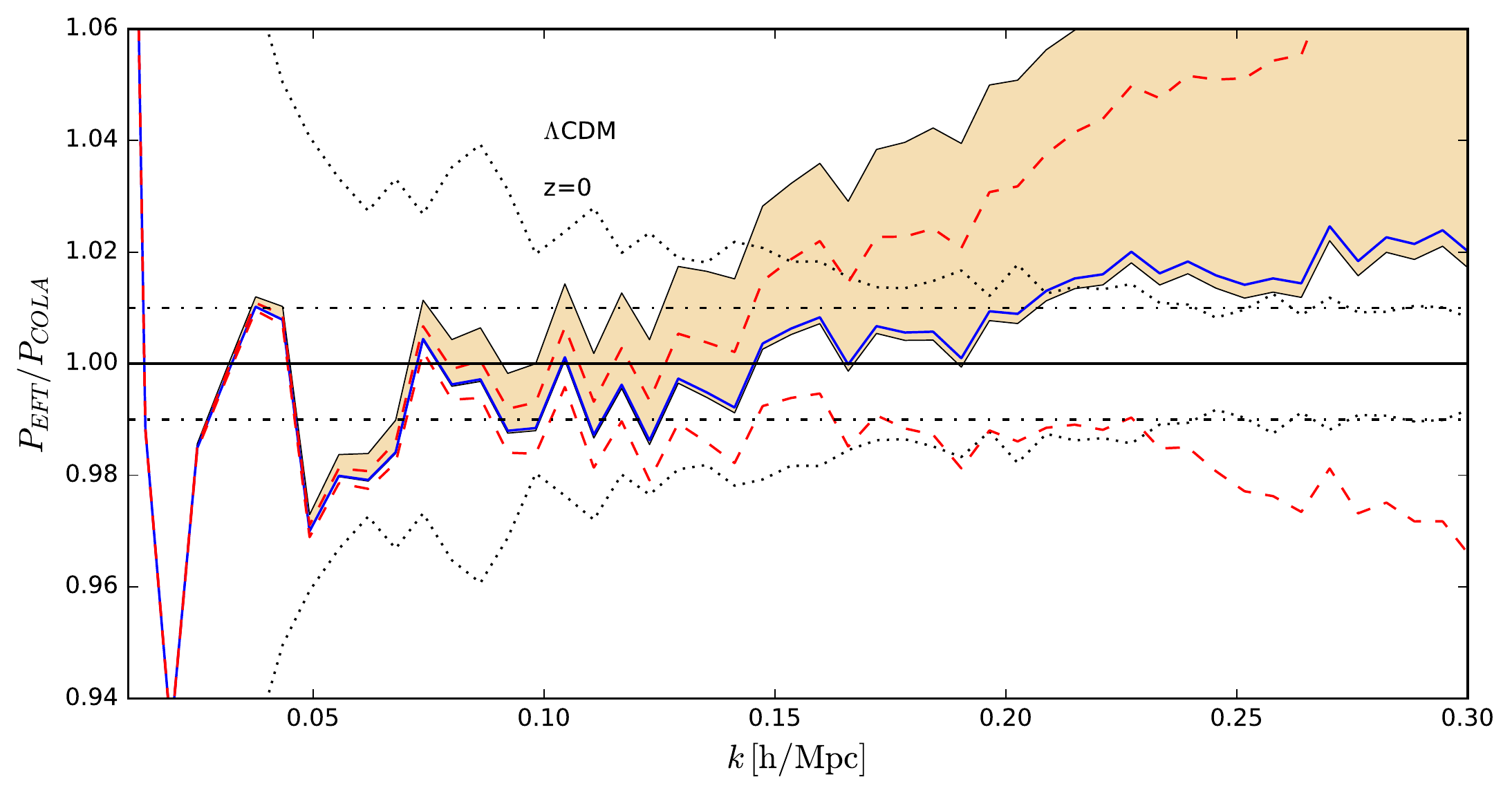}} 
  \caption[CONVERGENCE ]{ \footnotesize In this plot we compare two choices for the theoretical error at $z = 0$ for $\Lambda$CDM.  The yellow shaded region is due to the uncertainty in the fitting procedure, and is determined from the 1$\sigma$ errors for $\bar c_s^2$ at $0.75 k_{\rm fit}$.  The red dashed line is the estimated two-loop contribution using \eqn{scale11}.  As we can see here, the two are of approximately the same size.  At higher redshifts, the two-loop estimate gets smaller relative to the $0.75 k_{\rm fit}$ errors, so we are justified in using the $0.75 k_{\rm fit}$ errors in our analysis in this paper.}
\label{errors}
\end{figure}

We start with an estimate of the contribution from the first loop level not included in the computation.  To do this, we can consider the loop expression for a scaling universe, which is\footnote{We refer the reader to Eq.~(25) of \citee{Carrasco:2013mua} for more details on this computation.} 
\begin{align} 
P_{L\text{-loop}} / P_{11}\sim 2 \pi \left( k/ \knl \right)^{L( 3 + n)} \ .  \label{scale11} 
\end{align}
Approximate numerical values for $\knl$ and $n$ can be obtained by fitting the linear power spectrum to a scaling universe, i.e.
\be
P_{11}^{\rm scaling}(k) = \frac{ ( 2 \pi)^3}{\knl^3} \left( \frac{k}{\knl} \right)^n \  , 
\ee
over some range in $k$.  For example, at $z = 0$, when fitting the linear power spectrum to a scaling universe for $0.1 \unitsk < k < 0.2 \unitsk$, we find that $k_{\rm NL}^{\Lambda\text{CDM}} \approx 1.41 \unitsk$, $k_{\rm NL}^{\Omega_{\rm rc} = 0.438} \approx 1.25 \unitsk$, and the slope is a common $n \approx -1.61$.  In \figref{errors}, we show the estimate of the two-loop contribution for $\Lambda$CDM at $z=0$ as the red dashed line.  

This is to be compared with the uncertainty due to the fitting procedure described above.  To estimate this, we follow the procedure described in \cite{Foreman:2015lca}.   In order to capture the somewhat arbitrary nature of determining $\kfit$, the authors of \cite{Foreman:2015lca} decided to take the central value of $\bar c_s^2$ to be that determined at $\kfit$, and to take the uncertainty in $\bar c_s^2$ to be the variance in the $\chi^2$ fit at $0.75 \kfit$.  These two scales are shown as dotted and dot-dashed vertical lines respectively in \figrefss{kfits1}{kfits2}{kfits3}.  From these plots, one can see that the error bars at $0.75 \kfit$ are necessarily larger than the ones at $\kfit$, which indeed is guaranteed by the fitting procedure itself.  The error in the power spectrum induced by this uncertainty in $\bar c_s^2$ is shown as the yellow band in \figref{errors}.  As one can see, for the scales of interest, approximately $0.1 \unitsk \lesssim k \lesssim 0.2 \unitsk$, the uncertainty due to the fitting procedure is slightly larger than that due to the two-loop estimate.  In fact, at higher redshifts, the fitting procedure becomes even more the dominant source of error.  Thus, we come to the same conclusion as \cite{Foreman:2015lca}, and we choose to use the uncertainty due to the fitting procedure as the theoretical error quoted in this work.

\renewcommand{\bibname}{References}
 \bibliographystyle{utphys}
\bibliography{mybib,EFT_DE_biblio3}{}
\end{document}